\pdfoutput=1
\documentclass[11pt,twoside,a4paper,cmspaper,final,collab]{cms-tdr}
\def\svnVersion{7583fc3}\def\svnDate{2024/03/26}\def\cmsCernNoTag{CERN-EP-2023-223}\def\cmsCernDate{\today}\def\cmsMessage{Published in Physics Letters B as \href{http://dx.doi.org/10.1016/j.physletb.2024.138582}{\doi{10.1016/j.physletb.2024.138582}.}}
\begin{document}\cmsNoteHeader{HIG-22-003}

\newcommand{\sqrts}{\ensuremath{\sqrt{s}=13\TeV}\xspace}
\newcommand{\mhyp}{\ensuremath{m_{\Pa,\text{hyp}}}\xspace}
\newcommand{\ma}{\ensuremath{m_{\Pa}}\xspace}
\newcommand{\mllgg}{\ensuremath{m_{\ell\ell\PGg\PGg}}\xspace}
\newcommand{\mll}{\ensuremath{m_{\Pell\Pell}}\xspace}
\newcommand{\cZh}{\ensuremath{C_{\PZ\PH}^\text{eff}/\Lambda}\xspace}

\ifthenelse{\boolean{cms@external}}{\providecommand{\cmsLeft}{upper\xspace}}{\providecommand{\cmsLeft}{left\xspace}}
\ifthenelse{\boolean{cms@external}}{\providecommand{\cmsRight}{lower\xspace}}{\providecommand{\cmsRight}{right\xspace}}
\ifthenelse{\boolean{cms@external}}{\providecommand{\cmsTable}[1]{\resizebox{\columnwidth}{!}{#1}}}{\providecommand{\cmsTable}[1]{#1}}

\cmsNoteHeader{HIG-22-003}

\title{Search for an exotic decay of the Higgs boson into a \texorpdfstring{$\PZ$}{Z} boson and a pseudoscalar particle in proton-proton collisions at \texorpdfstring{$\sqrt{s} = 13\TeV$}{sqrt(s) = 13 TeV}}

\date{\today}

\abstract{
   A search for an exotic decay of the Higgs boson to a \PZ boson and a light pseudoscalar particle ($\Pa$), decaying to a pair of leptons and a pair of photons, respectively, is presented. The search is based on proton-proton collision data at a center-of-mass energy of \texorpdfstring{$\sqrt{s} = 13\TeV$}{sqrt(s) = 13 TeV}, collected with the CMS detector at the LHC and corresponding to an integrated luminosity of 138\fbinv. The analysis probes pseudoscalar masses \ma between 1 and 30\GeV, leading to two pairs of well-isolated leptons and photons. Upper limits at $95\%$ confidence level are set on the Higgs boson production cross section times its branching fraction to two leptons and two photons. The observed (expected) limits are in the range of 1.1--17.8 (1.7--17.9)\unit{fb} within the probed $\ma$ interval. An excess of data above the expected standard model background with a local (global) significance of 2.6 (1.3) standard deviations is observed for a mass hypothesis of $\ma=3\GeV$. Limits on models involving axion-like particles, formulated as an effective field theory, are also reported.
}

\hypersetup{%
   pdfauthor={CMS Collaboration},%
   pdftitle={Search for an exotic decay of the Higgs boson into a Z boson and a pseudoscalar particle in proton-proton collisions at sqrt{s} = 13 TeV},%
   pdfsubject={CMS},%
   pdfkeywords={CMS, axion like particles, Higgs}} 

\maketitle 

\section{Introduction}\label{sec:intro}

Following the discovery of the Higgs (\PH) boson by the ATLAS and CMS 
Collaborations~\cite{paper:Aad:2012,paper:Chatrchyan:2012,Chatrchyan:2013lba} at the CERN LHC, 
a thorough program of precision measurements~\cite{ATLAS:2022vkf,CMS:2022dwd} was carried out to uncover possible 
deviations from the predictions of the standard model (SM) of particle physics and to decipher 
the nature of the scalar sector. In particular, deviations in the \PH boson decay
width or the observation of exotic decay modes would constitute evidence of beyond-the-SM (BSM) physics. 

Axion-like particles, referred here as ALPs ($\Pa$), were originally proposed to address the strong $CP$ problem~\cite{PhysRevLett.38.1440}.
Recently, it was shown that ALPs could explain the observed anomaly in the magnetic moment of the muon~\cite{Buen-Abad:2021fwq}. Theoretical 
overviews of ALP models can be found in Refs.~\cite{Georgi:1986df,Bauer:2017ris}. 
The models are formulated as an effective field theory 
of ALPs coupled to various SM particles. In particular, these models allow couplings amongst the \PH boson, the \PZ boson, and the ALP fields, where the \PZ boson is found to be longitudinally polarized~\cite{Bauer:2016ydr}. 
Several searches involving ALPs, targeting different processes and final states, have been performed by the 
ATLAS and CMS Collaborations~\cite{atlas_ALP_1, atlas_ALP_2, atlas_ALP_3, atlas_ALP_4, ATLAS:2020pcy, ATLAS:2023zfc, CMS_ALP_1, CMS_ALP_2, CMS:2021pcy}. 
In particular, the search for $\PH \to \Pa \Pa \to \PGg\PGg\PGg\PGg$ has probed the existence of pseudoscalars with masses as low as 0.1\GeV~\cite{CMS_ALP_2}.

\begin{figure}[hbtp]
  \centering
    \includegraphics[width=0.49\textwidth]{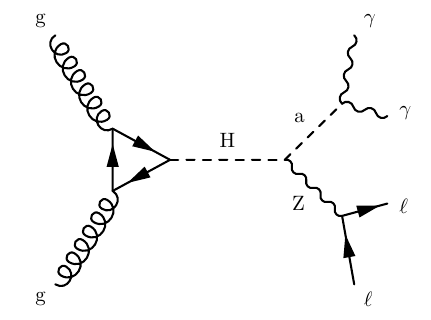}
    \caption{Feynman diagram for a BSM decay of the \PH boson into a \PZ boson and a light pseudoscalar boson, subsequently decaying to two leptons ($\Pell = \Pe, \PGm$) and two photons, respectively.}
    \label{fig:feyman}
\end{figure}

In this Letter, we report a search for an exotic decay of the \PH boson to an ALP and a \PZ boson, where the ALP decays
 to a pair of photons, and the \PZ boson decays to electrons or muons. 
The dominant Feynman diagram contributing to this process is shown in Fig.~\ref{fig:feyman}. Such a final state, 
with two charged leptons and two photons, results in an experimental signature that has low cross section in the SM~\cite{PhysRevD.71.017304}, and provides 
a complementary channel for the search for ALPs. It is the first search of this type at the LHC.

Assuming that the narrow-width approximation is valid for decays of the ALP and that the \PZ boson is on-shell, 
only the mass range $\ma < m_\PH - m_\PZ \approx 34\GeV$
is kinematically accessible to the $\PH \to \PZ \Pa$ decay, where $m_{\PH}$ and $m_{\PZ}$ are the $\PH$ and \PZ boson masses, respectively.
For $\ma$ below 1\GeV, the two photons originating from the ALP decay cannot be separated anymore in the detector~\cite{CMS:2022wjj}. 
Consequently, in this analysis, the mass range of $1 < \ma < 30\GeV$ is considered. 
The tabulated results are provided as HepData records~\cite{hepdata}.

\section{The CMS detector}\label{sec:detector}

The central feature of the CMS apparatus is a superconducting solenoid of 6\unit{m} internal diameter, 
providing a 3.8\unit{T} magnetic field. Within the solenoid volume are a silicon pixel and strip 
tracker, a lead tungstate crystal electromagnetic calorimeter (ECAL), and a brass and scintillator 
hadron calorimeter (HCAL), each composed of a barrel and two endcap sections. Forward calorimeters 
extend the pseudorapidity ($\eta$) coverage provided by the barrel and endcap detectors. Muons are detected 
in gas-ionization chambers embedded in the steel flux-return yoke outside the solenoid. A more 
detailed description of the CMS detector, together with a definition of the coordinate system used 
and the relevant kinematic variables, can be found in Ref.~\cite{Chatrchyan:2008zzk}.

Events of interest are selected using a two-tiered trigger system.
The first level, composed of custom hardware processors, uses information from the calorimeters 
and muon detectors to select events at a rate of around 100\unit{kHz} within a fixed latency of 4\mus~\cite{CMS:2020cmk}. 
The second level, known as the high-level trigger, consists of a farm of processors running a version of 
the full event reconstruction software optimized for fast processing, and reduces the event rate to around 1\unit{kHz} 
before data storage~\cite{Khachatryan:2016bia}.

\section{Data and simulated samples}\label{sec:datamc}

The proton-proton ($\Pp\Pp$) collision data at \sqrts used in this analysis were collected in 2016--2018 and 
correspond to a total integrated luminosity of 138\fbinv. 

Signal samples for the physical processes $\Pp\Pp \to \PH \to \PZ \Pa \to \Pell\Pell\PGg\PGg$ (where $\Pell = \Pe, \PGm$, and leptonic $\Pgt$ decays are included in the signal) 
are generated at leading order (LO) for $\ma$ range of
1--10\GeV in 1\GeV steps, and 10--30\GeV in 5\GeV steps, 
using the \MGvATNLO 2.2.2 (2.4.2)~\cite{MADGRAPH5,Alwall:2007fs,Frederix:2012ps}
generator for analyzing data collected in 2016 (2017--2018). The \PH boson mass is assumed to be 125\GeV, 
and only the dominant production mode of the \PH boson, the gluon fusion, is considered. 
Other production modes were evaluated, but they were found to have comparable signal selection efficiencies relative to the gluon fusion mode, 
with no more than a 10\% relative difference. Consequently, since the final results are normalized to the \PH boson inclusive production cross section, 
differences in kinematics between the different production modes have a negligible impact ($<$1\%) on the predicted signal normalization.

The predominant background in this search is the SM Drell--Yan production of $\PZ+\text{jets}$, in which jets are misidentified 
as photons. As in Refs.~\cite{Khachatryan:2014ira,Sirunyan:2018ouh}, this background contribution is modeled entirely from data to 
extract the results. 
For the analysis selection optimization, simulated Drell--Yan events are used. They are generated at next to leading order with the \MGvATNLO 
2.2.2 (2.4.2)~\cite{MADGRAPH5,Alwall:2007fs,Frederix:2012ps}
generator for the analysis of 2016 (2017--2018) data. 

The set of parton distribution functions (PDFs) NNPDF3.0~\cite{Ball:2014uwa} (NNPDF3.1~\cite{Ball:2017nwa}) is used for the 2016 (2017--2018) simulation.
Parton showering and hadronization are simulated using the \PYTHIA~8.230 generator~\cite{Sjostrand:2014zea}
with the CP5 underlying event tune for the simulation of all three years~\cite{Khachatryan:2015pea,Sirunyan:2019dfx}.
The response of the CMS detector is modeled using the \GEANTfour toolkit~\cite{AGOSTINELLI2003250,1610988}.
Additional interactions in the same or adjacent bunch crossings (pileup) are modeled by superimposing simulated minimum bias collisions on the hard-scattering interaction, with the multiplicity matching that observed in data.

\section{Event reconstruction}\label{sec:reco}

The primary vertex is taken to be the vertex corresponding to the hardest scattering in the event, evaluated using tracking information alone~\cite{CMS-TDR-15-02}.

The particle-flow (PF) algorithm~\cite{CMS-PRF-14-001} aims to reconstruct and identify each individual particle in an event (PF candidate), 
with an optimized combination of information from the various elements of the CMS detector. 
The energy of photons is obtained from the ECAL energy deposition in a supercluster,
combining deposits from the photon and the conversions, 
produced by the tracker material upstream of the ECAL detector~\cite{CMS:2020uim}. 
The energy of the electrons is estimated from a combination of the electron momentum at the primary interaction vertex 
as determined by the tracker, 
the energy of the corresponding ECAL cluster, and the energy sum of all bremsstrahlung photons spatially compatible 
with the origin of the electron track. 
A multivariate regression technique is employed to correct the photon and electron energies measured in the ECAL. These procedures are described in Ref.~\cite{CMS:2020uim}. 
The energy of muons is obtained from the curvature of the corresponding track. The energy of the charged hadrons is determined from 
a combination of their momentum measured in the tracker and the matching ECAL and HCAL energy deposits, corrected for the response function of 
the calorimeters to hadronic showers. Finally, the energy of neutral hadrons is obtained from the corresponding corrected ECAL and HCAL energies.

Electrons with the transverse momentum $\pt > 7\GeV$ and $\abs{\eta} < 2.5$ are identified with a multivariate discriminant, which 
is constructed using variables related to the bremsstrahlung along the electron trajectory, ECAL energy measurements, electromagnetic showers, missing pixel detector hits, 
and the photon conversion vertex fit probability.
The momentum resolution for electrons with $\pt \approx 45$\GeV from $\PZ \to \Pe \Pe$ decays, where $\Pe$ refers to $\Pep$ and $\Pem$, is in the range 1.7--4.5\%.
For electrons with $\pt \lesssim 30\GeV$, the momentum resolution is in the range 2.1--6.1\%~\cite{CMS:2020uim}.
It is generally better in the barrel than in the endcaps, and also depends on the bremsstrahlung energy emitted by the electron as it traverses the material in front of the ECAL.

Muons are measured in the range $\abs{\eta} < 2.4$, with detection planes made using three technologies: drift tubes, 
cathode strip chambers, and resistive-plate chambers. The efficiency of the single-muon trigger, which has a 20\GeV \pt threshold, exceeds 90\% over the full $\eta$ range, and the 
efficiency to reconstruct and identify muons is greater than 96\%. Matching muons to tracks measured in the silicon tracker results in a 
relative \pt resolution, for muons with \pt up to 100\GeV, of 1\% in the barrel and 3\% in the endcaps~\cite{Sirunyan:2018fpa}.

To reduce the contributions from leptons arising from hadron decays within jets, a requirement is imposed on each lepton candidate using their 
relative isolation, defined as:
\ifthenelse{\boolean{cms@external}}{
\begin{multline}
I^{\Pell} = \Biggl(\sum \PT^{\text{charged}}\\
+\max\Bigl[0,\sum \PT^{\text{neutral}}
+\sum \PT^{\PGg}-\PT^{\text{PU}}(\Pell)\Bigr]\Biggr)\Bigg/\PT^{\Pell}
\end{multline}
}
{
\begin{equation}
I^{\Pell} = \Biggl(\sum \PT^{\text{charged}}
+\max\Bigl[0,\sum \PT^{\text{neutral}}
+\sum \PT^{\PGg}-\PT^{\text{PU}}(\Pell)\Bigr]\Biggr)\Bigg/\PT^{\Pell}
\end{equation}
}
where the sums are over the PF candidates within an angular cone of radius $\Delta R < 0.3$ (where $\Delta R = \sqrt{(\Delta\eta)^2+(\Delta\phi)^2}$ and $\phi$ is the azimuthal angle in radians), $\PT^{\text{i}}$ represents the \pt of individual 
charged or neutral hadrons, photons, or pileup~\cite{Sirunyan:2017exp}. 
For electrons, the relative isolation variable, $I^{\Pe}$, is included in the multivariate discriminant. 
For muons, the relative isolation is required to be $I^{\mu} < 0.35$.
In addition, the three-dimensional impact parameter of electrons and muons is required to be consistent with the primary collision vertex. 

An algorithm is utilized to account for final-state radiation (FSR) from leptons. Photons reconstructed with the PF algorithm are considered as 
FSR candidates if they satisfy the requirement $\PT^{\PGg} > 2\GeV$ and $I^{\PGg} < 1.8$, where $I^{\PGg}$ is calculated similarly to the lepton 
isolation variable. Each FSR candidate is then assigned to the closest lepton in the event. The candidates are further required to have 
$\Delta R(\PGg,\Pell)/\left(\PT^{\PGg}\right)^2 < 0.012\GeV^{-2}$ and $\Delta R(\PGg,\Pell) < 0.5$. 
These FSR photon candidates are excluded from the calculation of the lepton isolation variables. 
The Higgs boson mass resolution is improved by about 1 -- 4\% with this FSR recovery algorithm.

Lepton reconstruction and selection efficiencies are measured in data with the ``tag-and-probe'' technique with an inclusive sample of \PZ boson events~\cite{Khachatryan:2010xn}. The differences 
between the efficiencies in data and simulation are observed to be around 1--4\%, depending on the $\PT$ and $\eta$ of the lepton considered. 
They are used to correct lepton efficiencies in simulation.

Prompt photons are distinguished from jets using a cut-based technique~\cite{Khachatryan:2014ira} that uses information 
related to the photon isolation variables, the ratio of the energy in the HCAL behind an electromagnetic supercluster to the supercluster energy, and the transverse width of the electromagnetic shower ($\sigma_{i\eta i\eta}$). Because the two photons  
are collimated for low \ma, their $\sigma_{i\eta i\eta}$ and photon isolation in the ECAL, 
which reflect the relative contribution of energy from nearby photons in the ECAL, depend on each other. 
In this analysis, $\sigma_{i\eta i\eta}$ and the photon isolation in the ECAL are removed from the identification requirement. 
Photons are required to lie in the geometrical region $\abs{\eta} < 2.5$ and to have $\pt > 10\GeV$. 
The photon identification efficiency is measured using $\PZ\to\Pe\Pe$ events with electrons reconstructed as photons, using the tag-and-probe technique.

\section{Event selection optimization}\label{sec:selection}

A set of event requirements is defined to maximize the sensitivity to a potential signal. The leading muon (electron) is required 
to have $\pt > 20$ (25)\GeV and the subleading one $\pt > 10$ (15)\GeV, while each photon is required to have $\pt > 10\GeV$.

For each event, \PZ boson candidates are formed by considering all opposite-sign, same-flavor lepton pairs with 
the dilepton invariant mass $\mll$ closest to the nominal \PZ boson mass, 
while ALP candidates are built from selected photons with the highest leading and subleading $\pt$. For each dilepton and diphoton 
candidate, a $\PZ \Pa$ candidate is constructed. The $\PZ \Pa$ candidates are required to satisfy $\mll > 50\GeV$, 
$95 < \mllgg < 180\GeV$, and $\Delta R(\Pell,\PGg) > 0.4$. 
A loose dilepton invariant mass requirement, $\mll > 50\GeV$, is applied in order not to introduce any mass peak sculpting in the $\mllgg$ mass distribution for high masses of ALP candidates.

A boosted decision tree (BDT) event classifier is trained to separate the signal from the background. 
In order to make the classifier output uniform and sensitive to the full \ma range considered in this search, 
the training sample is parameterized as a function of \ma~\cite{Baldi_2016}. In this approach, 
a parameter equal to the hypothesized pseudoscalar mass $\mhyp$ is provided as one of the inputs to the training.
The parameterized classifier requires only one single training and is able to provide interpolation to intermediate \ma  hypotheses 
that were not used during the training. Simulations corresponding to various \ma hypotheses are combined and treated as 
a common signal in the training. The value of $\mhyp$ is set equal to the mass hypothesis for the signal simulation. 
For the background, simulated Drell--Yan events are used in the training and the parameter $\mhyp$ is randomly distributed as a flat 
function in the \ma range.
The variables used to train the BDT are chosen such that the value of \mllgg cannot be inferred from the inputs. 
For this purpose, the correlations between the input variables and \mllgg are checked, and the variables that have a strong correlation with \mllgg are removed.
Finally, the following discriminating variables are used as input to the BDT:
\begin{itemize}
    \item $\pt(\PGg1)$ and $\pt(\PGg2)$, where $\PGg1$ and $\PGg2$ are the leading and subleading photons, respectively; 
    \item $R_9(\PGg1)$ and $R_9(\PGg2)$: the energy sum of the $3{\times}3$ crystal array centered around the most energetic crystal in the supercluster, divided by the energy of the supercluster;
    \item $\sigma_{i\eta i\eta}(\PGg1)$ and $\sigma_{i\eta i\eta}(\PGg2)$: the second moment of the log-weighted distribution of crystal energies in $\eta$, calculated in the $5{\times}5$ matrix around the most energetic crystal in the supercluster and rescaled to units of crystal size;
    \item $I_{\PGg}(\PGg1)$ and $I_{\PGg}(\PGg2)$: the isolation variable obtained by summing the \pt of photons inside an isolation cone of $\Delta R = 0.3$ with respect to the photon direction, while the impact of another selected photon is also included;
    \item $I_{\PGg,\Pa}$, the isolation variable obtained by summing the \pt of photons inside an isolation cone of $\Delta R = 0.3$ with respect to the direction of the ALP candidate;
    \item the angular separation between the \PZ boson and the diphoton pair, $\Delta R(\PZ,\Pa)$;
    \item the angular separation between the two photons, $\Delta R(\PGg_{1},\PGg_{2})$;
    \item the angular separation between the leading photon and the \PZ boson, $\Delta R (\PGg_{1},\PZ)$;
    \item the ALP candidate's \pt divided by $\mllgg$;
    \item the \PH boson candidate's \pt;
    \item the difference between the invariant masses of the ALP candidate and the \mhyp parameter divided by \mllgg, $(\ma-\mhyp)/\mllgg$. 
\end{itemize}

Because the leptons have been selected with tight requirements, 
lepton identification criteria are not included as discriminating variables.
In addition, the fact that angular variables have low ranking among the BDT input variables makes the difference between pseudoscalar and scalar assumptions negligible.
The simulated signal and background data sets from the different data-taking years are scaled by the corresponding integrated luminosity and 
combined into a large training sample. Additionally, signal and background events are divided into independent training 
and testing samples to confirm the absence of overtraining.

\begin{figure*}[hbt!]
    \centering
      \includegraphics[width=0.49\textwidth]{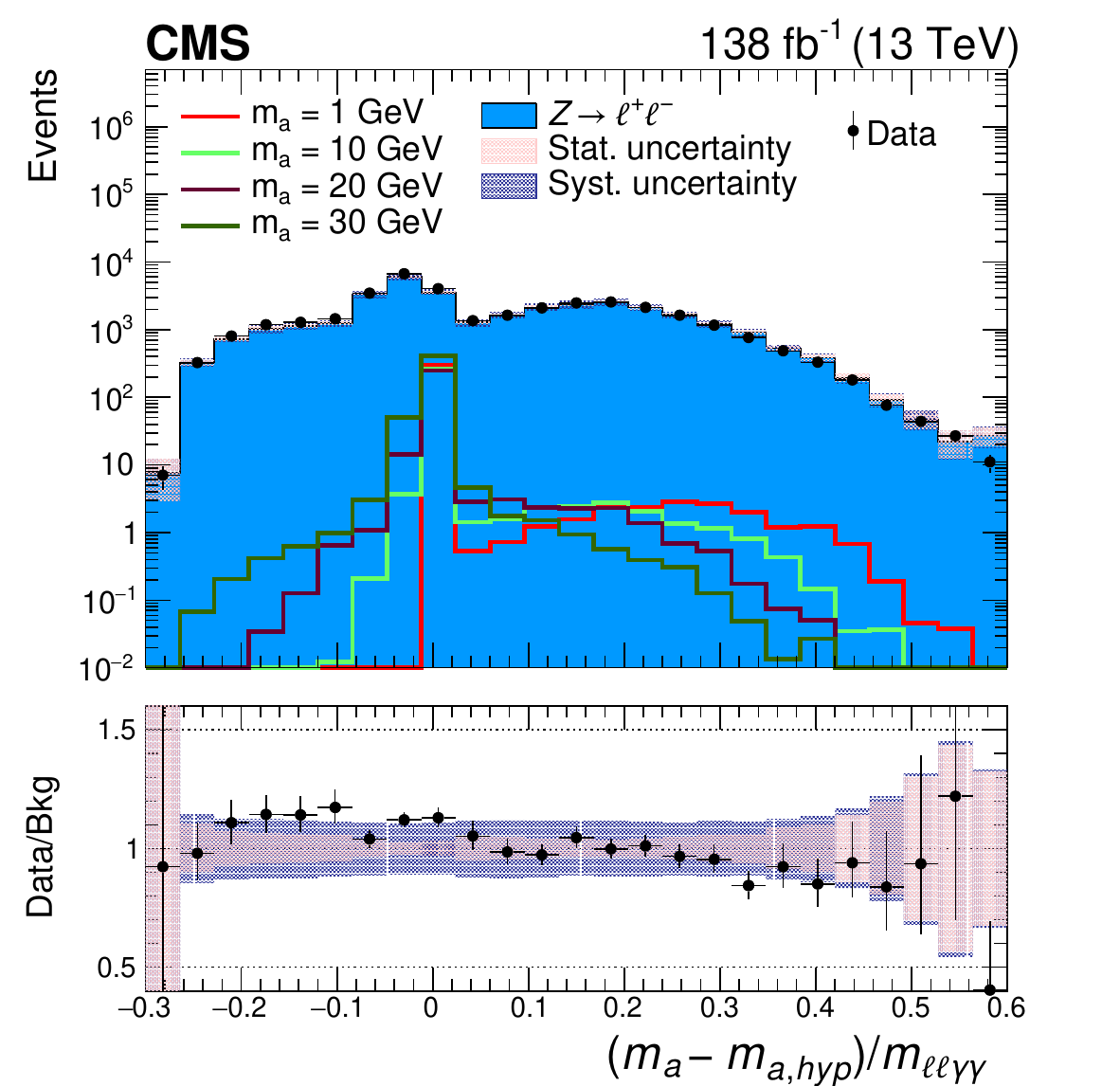}
      \includegraphics[width=0.49\textwidth]{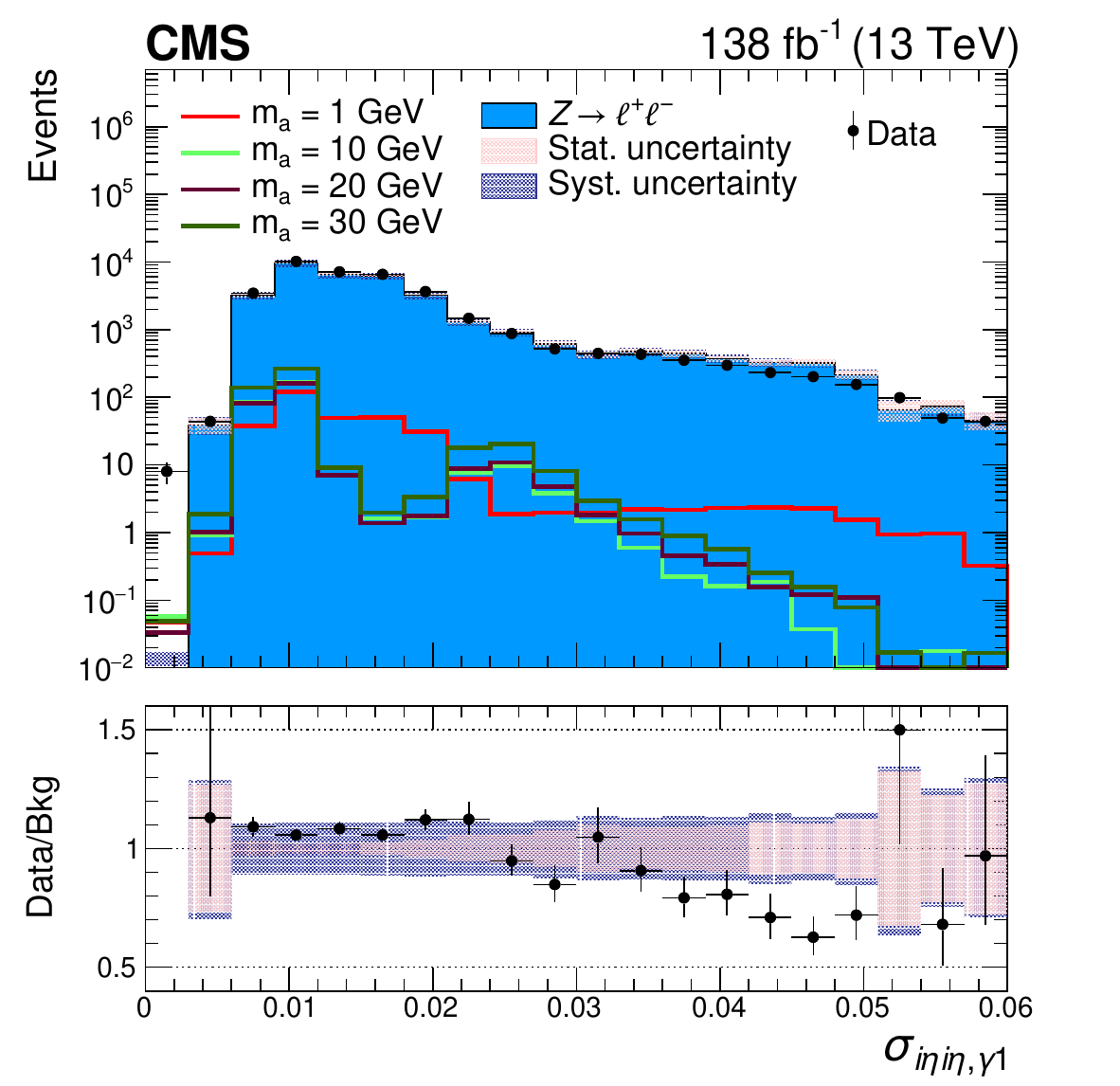}
  
      \includegraphics[width=0.49\textwidth]{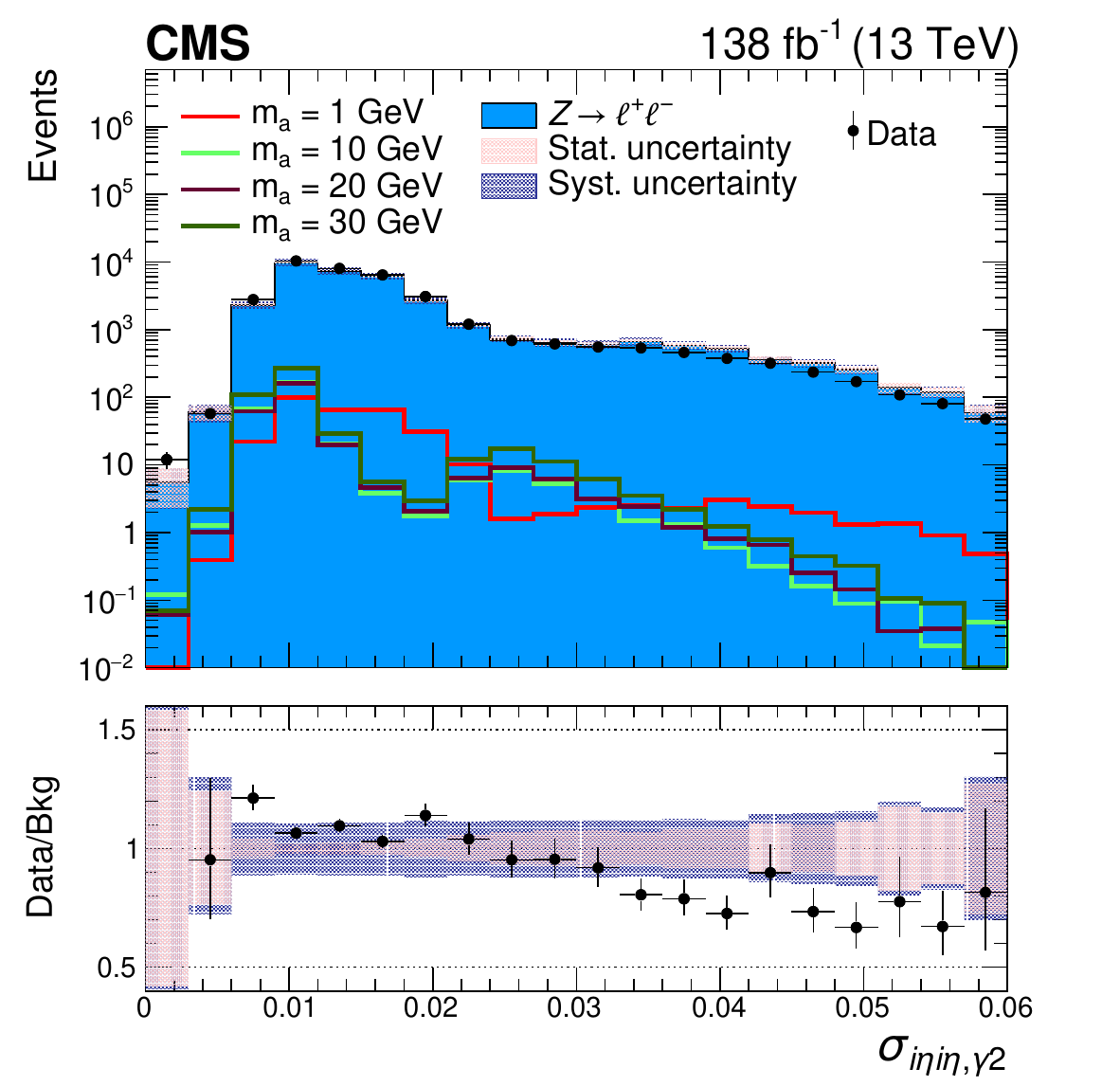}
      \includegraphics[width=0.49\textwidth]{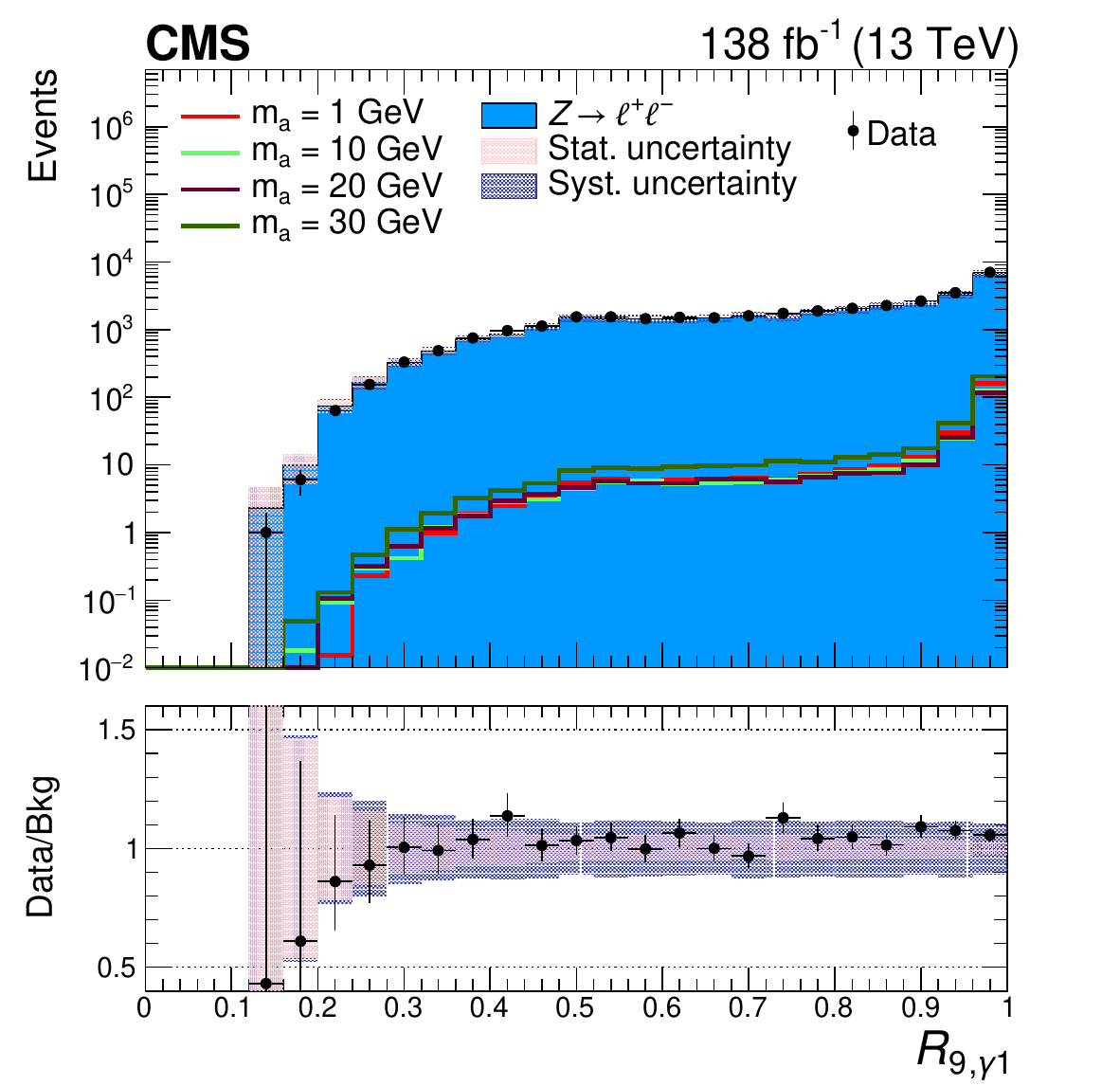}
      \caption{Distributions of the four most discriminating variables used as input to the BDT: $(\ma-\mhyp)/\mllgg$ (upper left), leading photon's $\sigma_{i\eta i\eta}$ (upper right), subleading photon's $\sigma_{i\eta i\eta}$ (lower left), and leading photon's $R_9$ (lower right). The events pass the selection criteria described in Section~\ref{sec:selection}. The signal is scaled to a cross section of 0.1\unit{pb} and the background sample is normalized to an integrated luminosity of 138\fbinv. The systematic uncertainties included in the shaded band are related to the photon efficiency, lepton efficiency, and pileup modeling. The impact of the remaining disagreement between data and simulation is negligible.}
      \label{fig:var}
\end{figure*}

The four most discriminating variables, shown in Fig.~\ref{fig:var}, are $(\ma-\mhyp)/\mllgg$, the leading photon's $\sigma_{i\eta i\eta}$, the 
subleading photon's $\sigma_{i\eta i\eta}$, and the leading photon's $R_9$.
The distributions of the simulated background and the data are found to be in reasonable agreement. Since the background model used 
to extract the limits is estimated from data (see Section~\ref{sec:bkgmodel}), any remaining disagreement between data and simulated background does not bias the final results. 
The shape discrepancy between the data and simulated background is propagated to the final BDT selection, and the impact is checked to be negligible (${<}2\%$).

\begin{figure*}[hbt!]
    \centering
      \includegraphics[width=0.49\textwidth]{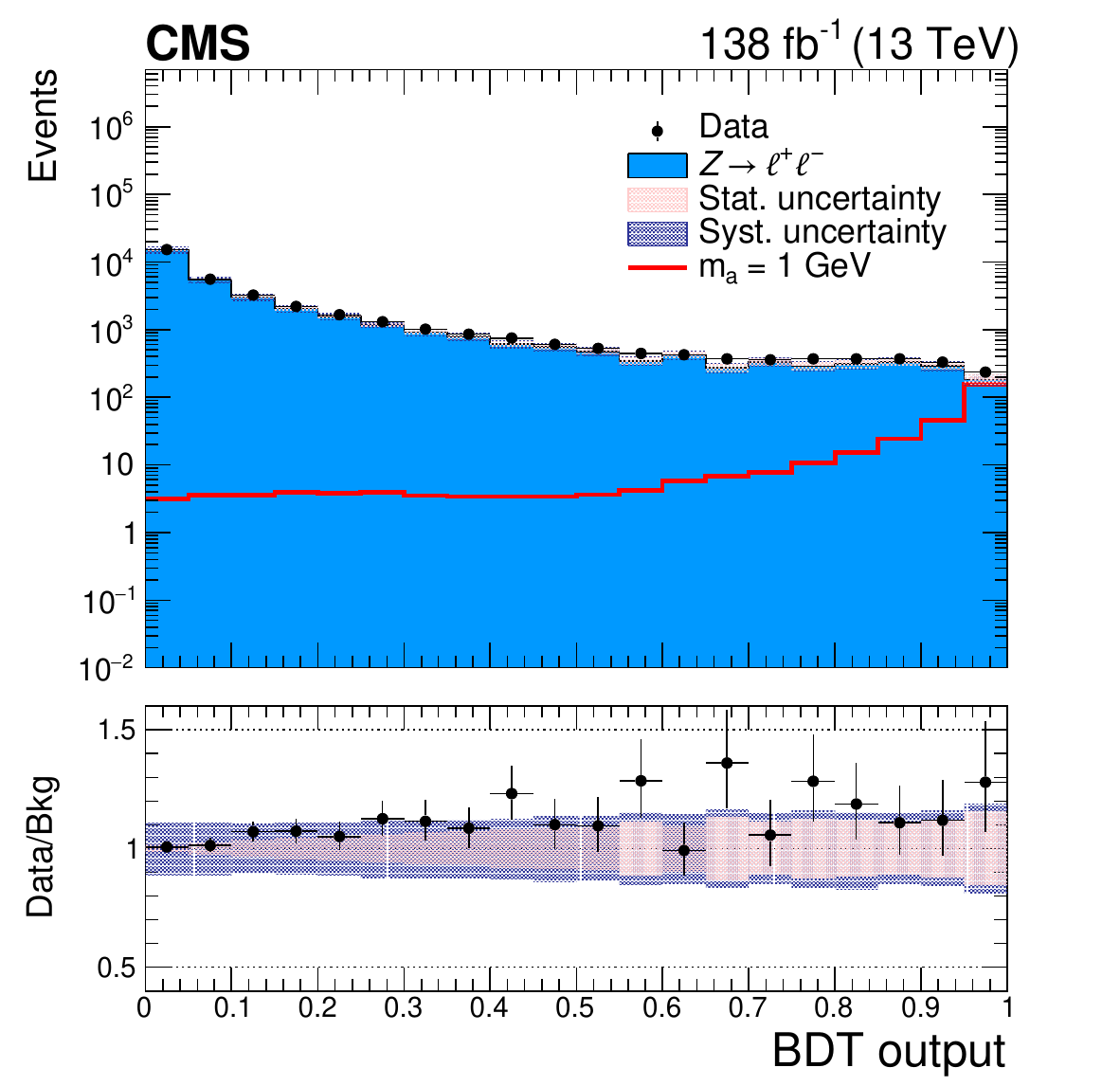}
      \includegraphics[width=0.49\textwidth]{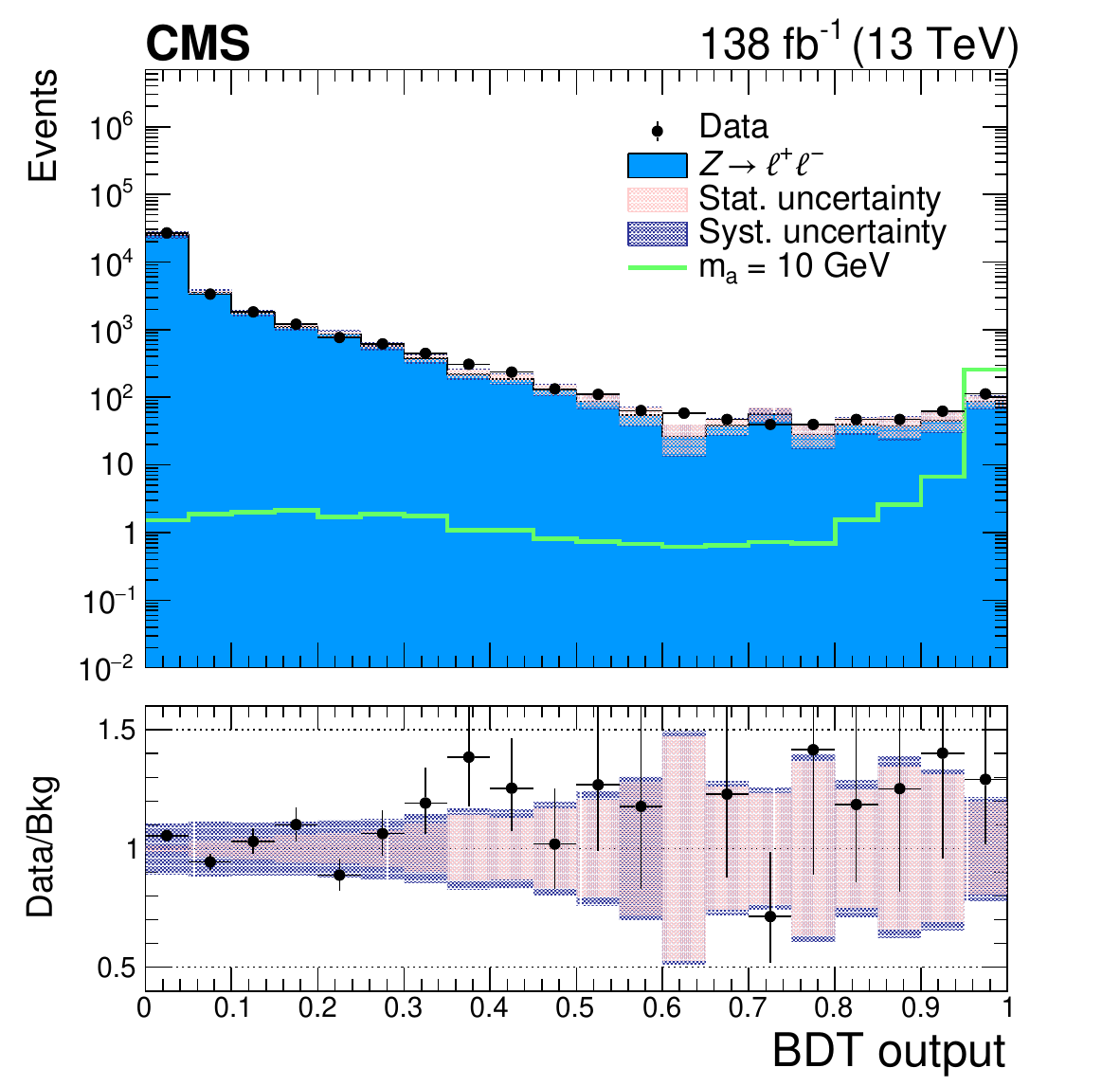}
  
      \includegraphics[width=0.49\textwidth]{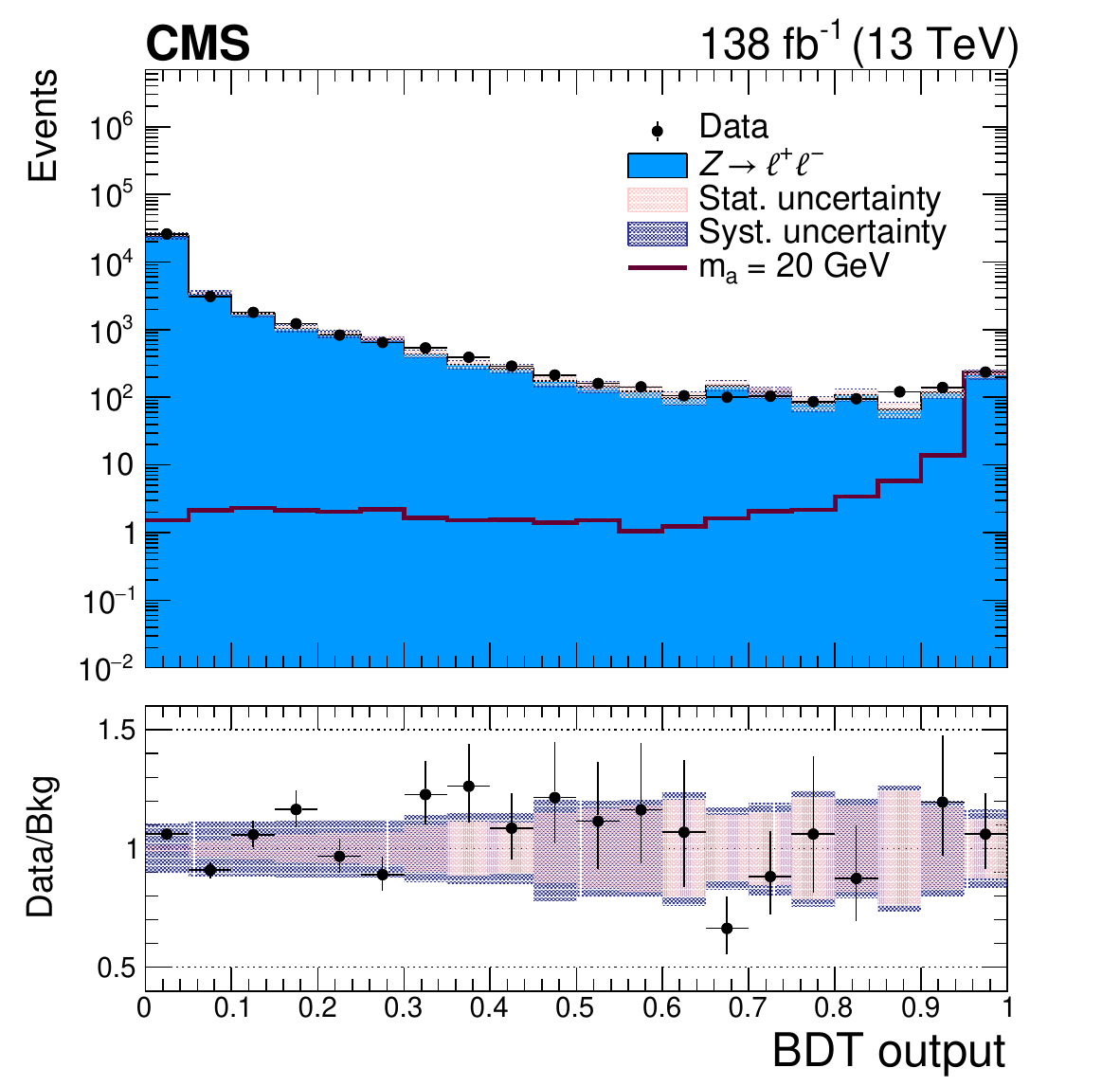}
      \includegraphics[width=0.49\textwidth]{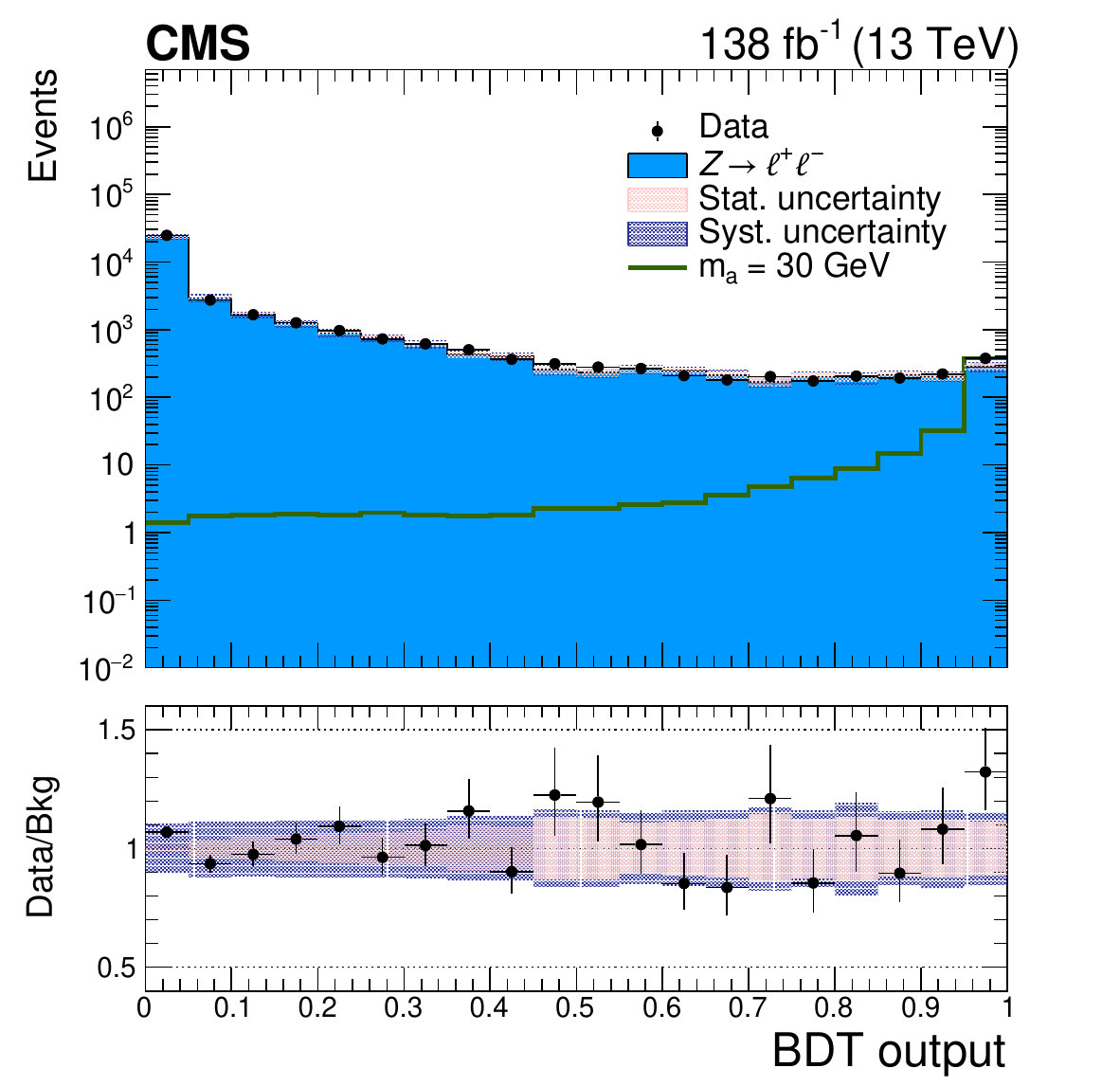}
      \caption{Distributions of the BDT output for $\ma = 1\GeV$ (upper left), 10\GeV (upper right), 20\GeV (lower left), and 30\GeV (lower right). The events pass the selection criteria described in Section~\ref{sec:selection}. The signal is scaled to a cross section of 0.1\unit{pb} and the background sample is normalized to an integrated luminosity of 138\fbinv.  The systematic uncertainties included in the shaded band are related to the photon efficiency, lepton efficiency, and pileup modeling.}
      \label{fig:bdtout}
\end{figure*}

A unique BDT output is obtained for each \ma hypothesis. Figure~\ref{fig:bdtout} shows the distributions of the BDT outputs 
for data and simulated events for $\ma = 1, 10, 20,$ and 30\GeV.

Events are categorized according to the output of the BDT. The BDT thresholds maximize the approximate mean 
significance (AMS)~\cite{Cowan:2010js} over all possible categories in the signal region ($115 <\mllgg< 135\GeV$), separately for each \ma.  
The AMS is defined as:
\begin{equation}
    \mathrm{AMS} = \sqrt{2\left[(S+B)\ln{(1+\frac{S}{B})} - S\right]}\label{con:AMS}.
\end{equation}

In Eq. (\ref{con:AMS}), $S$ and $B$ refer to the number of signal and background (Drell--Yan simulation) events in the signal region. 
In order to minimize the impact of statistical fluctuations, the output BDT distribution of the background is smoothed, using the super-smoothing technique~\cite{SmmothSuper,SmmothSuper2}. 
This optimization procedure is performed separately for each \ma hypothesis for up to two categories, based on the BDT score. 
An increase of less than 1\% in the AMS value is observed when increasing the number of categories beyond 1. 
Therefore, a single category based on the BDT output is considered. Table~\ref{tab:boundary} details the category boundaries, 
the related signal efficiencies, and the Drell--Yan background yields.

\begin{table}[ht!]
    \centering
      \topcaption{Minimum BDT output values used to define the signal region, with the associated signal efficiencies and background yields in the signal region. The statistical uncertainties are also shown.}
      \label{tab:boundary}
      \cmsTable{
      \begin{tabular}{ccccc} 
        $\ma$ (\GeV) & Min. BDT & Signal efficiency & Drell--Yan\\ 
        & output value & (\%) & background yields \\ \hline
       1 & 0.955 & $49 \pm 3.3$ & $83 \pm 27$ \\
       2 & 0.980 & $67 \pm 2.7$ & $26 \pm 10$ \\
       3 & 0.985 & $76 \pm 2.4$ & $7.9 \pm 4.9$ \\
       4 & 0.980 & $84 \pm 2.1$ & $5.1 \pm 4.5$ \\
       5 & 0.985 & $85 \pm 2.1$ & $5.1 \pm 3.9$ \\
       6 & 0.990 & $82 \pm 2.3$ & $2.5 \pm 2.2$ \\
       7 & 0.985 & $86 \pm 2.1$ & $5.3 \pm 4.0$ \\
       8 & 0.990 & $80 \pm 2.5$ & $11 \pm 4.8$ \\
       9 & 0.990 & $78 \pm 2.5$ & $16 \pm 5.6$ \\
       10 & 0.990 & $77 \pm 2.6$ & $11 \pm 4.7$ \\
       15 & 0.990 & $70 \pm 2.9$ & $13 \pm 5.2$ \\
       20 & 0.990 & $63 \pm 3.1$ & $18 \pm 6.1$ \\
       25 & 0.985 & $64 \pm 2.7$ & $37 \pm 11$ \\
       30 & 0.980 & $67 \pm 2.2$ & $44 \pm 13$ \\
      \end{tabular}}
\end{table}

\section{Statistical procedure}\label{sec:stat}

The statistical procedure to extract the results 
is identical to that described in Ref.~\cite{likelihoodRatio}. 
An unbinned maximum likelihood fit is performed on the \mllgg distribution in the mass range $95 < \mllgg < 180\GeV$ for each $\ma$ hypothesis, 
with an \ma granularity of 1\GeV in the range 1--30\GeV. 
A likelihood function is defined for each $\ma$ hypothesis using analytic models that describe the \mllgg distributions of 
signal and background processes. Nuisance parameters are included to account for the experimental and theoretical 
systematic uncertainties described in Section~\ref{sec:syst}.
The data collected in 2016--2018 are combined, with the electron and muon channels merged together.
The best-fit values and confidence intervals for the parameters of interest are estimated using a profile likelihood test statistic:
\begin{linenomath*}
    \begin{equation}
    \label{eq:likelihood}
        q(\vec{\alpha}) = - 2\ln \left( \frac{L(\vec{\alpha},\hat{\vec{\theta}}_{\vec{\alpha}})}{L(\hat{\vec{\alpha}},\hat{\vec{\theta}})} \right),
    \end{equation}
\end{linenomath*}
where $\hat{\vec{\alpha}}$ and $\hat{\vec{\theta}}$ describe the unconditional 
maximum likelihood estimates for the parameters of interest and the nuisance parameters, 
respectively, whereas $\hat{\vec{\theta}}_{\vec{\alpha}}$ corresponds to the conditional 
maximum likelihood estimate for fixed values of the parameters of interest, $\vec{\alpha}$. 
    
\section{Signal model}\label{sec:sigmodel}

The signal shape for the \mllgg distribution, for each nominal signal hypothesis, is constructed from simulation. 
After all the selection criteria are applied, a signal model is built for each \ma hypothesis 
and for each data-taking year. The \mllgg distribution is modeled with a sum of $n$ Gaussian functions ($n<5$). 
The electron and muon channels are treated separately. 
These signal models, scaled by the corresponding integrated luminosity, are summed together to construct the final signal model.

\begin{figure}[hbt!]
    \centering
      \includegraphics[width=0.49\textwidth]{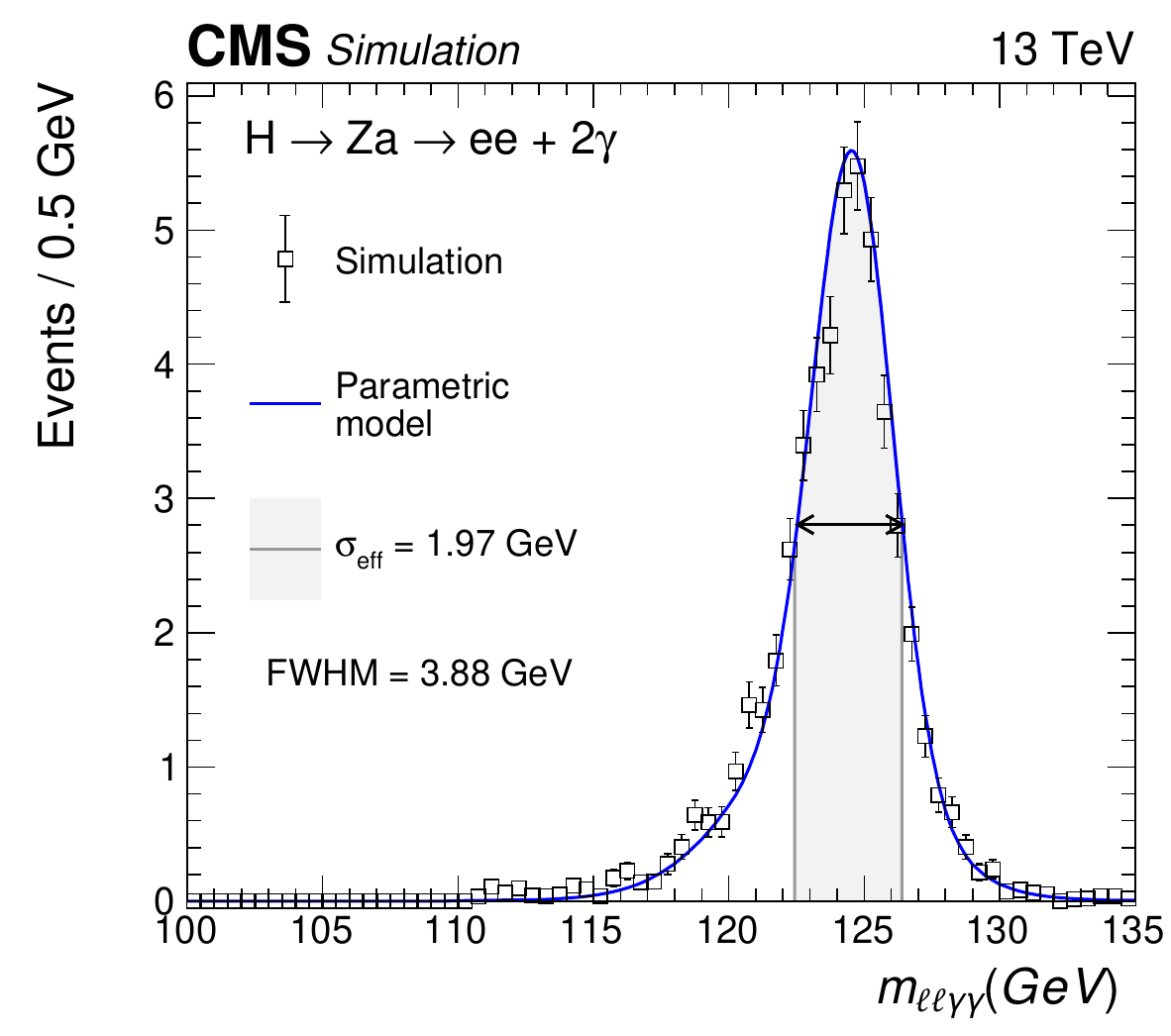}
      \includegraphics[width=0.49\textwidth]{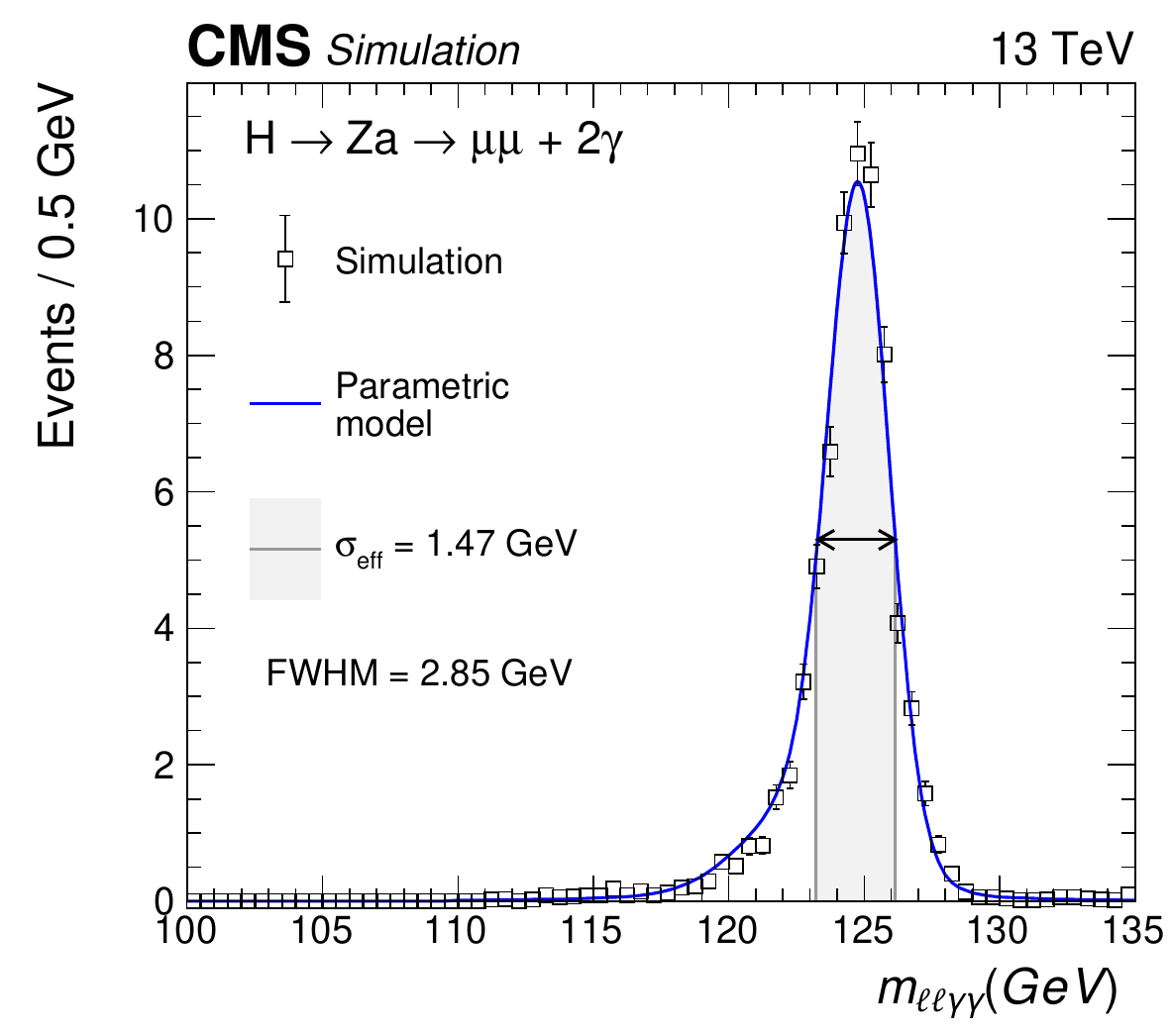}
      \caption{Fit to the simulated \mllgg distributions for a signal with $\ma=30\GeV$ in the electron (\cmsLeft) and muon (\cmsRight) channels for the year 2018.}
      \label{fig:sigModel_M30}
\end{figure}

The number of Gaussian functions used for the signal modeling is determined with an $F$-test \cite{ftest}. 
First, an F-test is performed to determine the order of the Gaussian fit, and then the parameters that fit the signal 
distribution best are extracted.
As an example, the signal models are shown in Fig.~\ref{fig:sigModel_M30} for $\ma=30\GeV$ in the electron and muon channels and for the year 2018. 
The full width at half maximum (FWHM) and the effective standard deviation ($\sigma_\text{eff}$), which is defined as half 
the width of the smallest interval containing 68\% of the \mllgg distribution, are also shown. 

\begin{figure}[hbt!]
  \centering
    \includegraphics[width=0.49\textwidth]{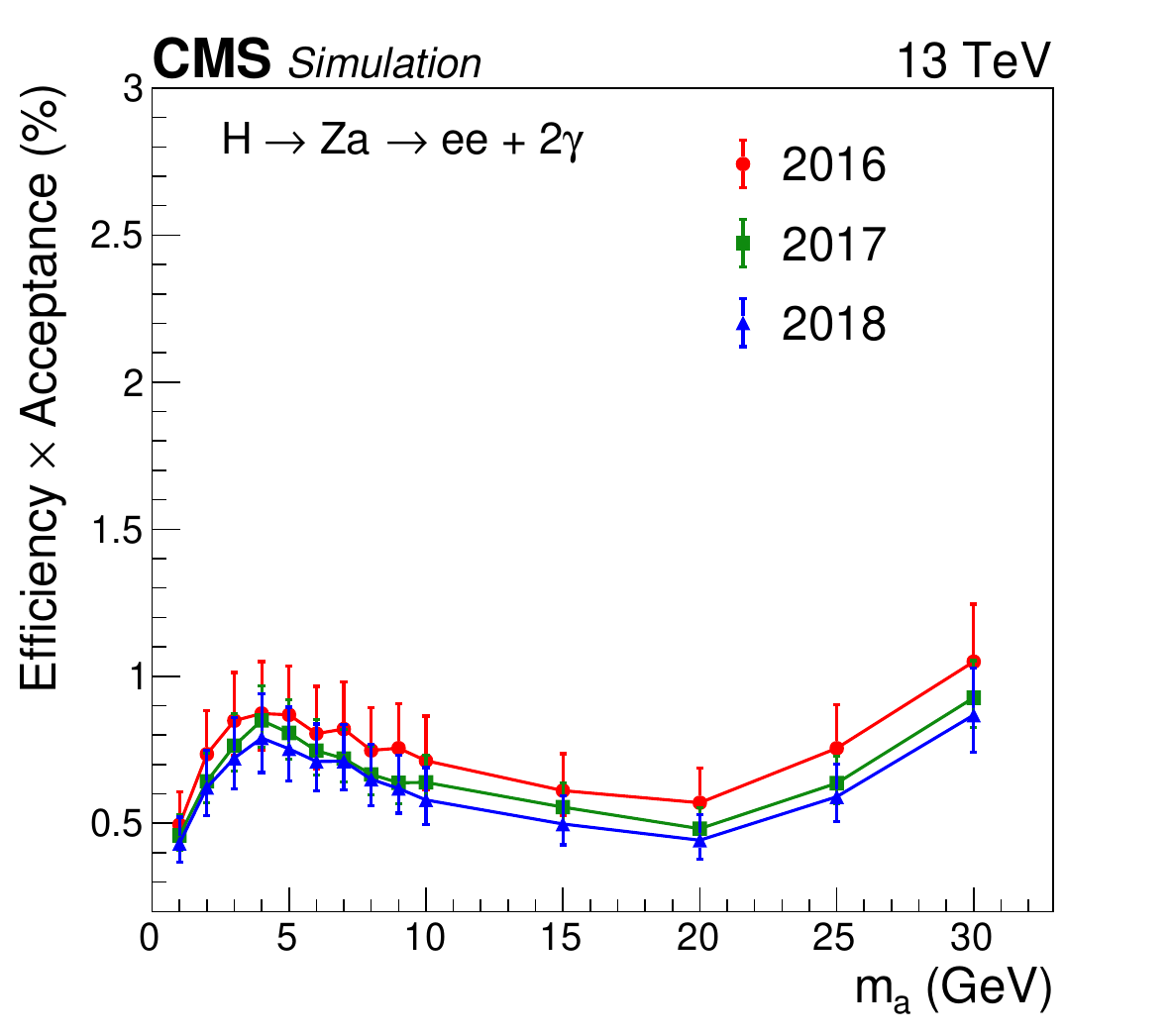}
    \includegraphics[width=0.49\textwidth]{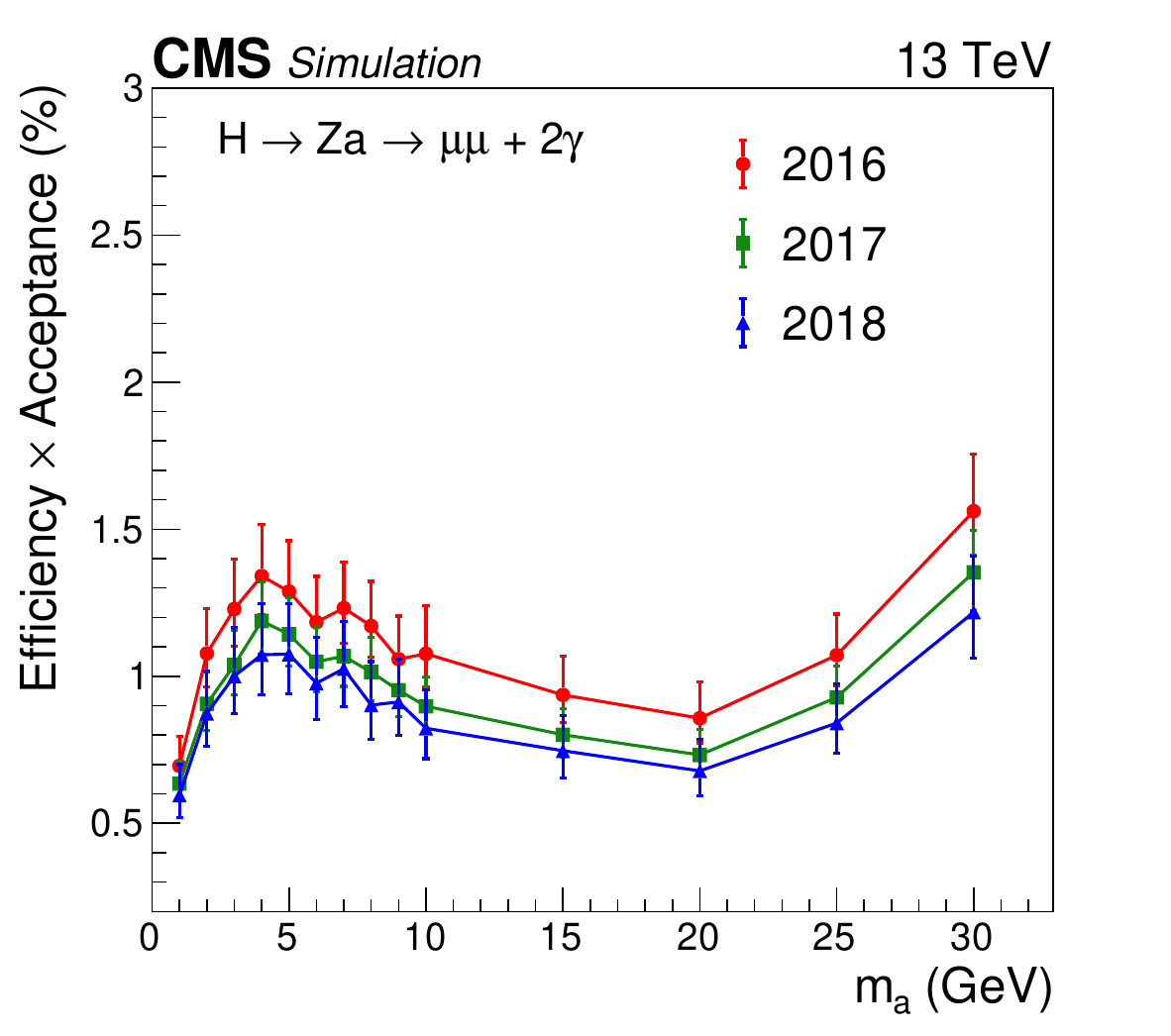}
    \caption{Product of detector efficiency and analysis acceptance for signal samples with various \ma values for the electron (\cmsLeft) and muon channel (\cmsRight). The error bars include statistical and systematic uncertainties. The photon efficiency, lepton efficiency, and pileup modeling uncertainties are taken into account for the systematic uncertainty.}
    \label{fig:eff}
\end{figure}

To build the signal models for the intermediate mass hypotheses in the range $10 <\ma <30\GeV$, two factors must be considered: 
the shape of the \mllgg distribution and its normalization. Since the shape of the \mllgg distribution does not significantly depend 
on $\ma$ in the interpolation range, only the normalization of the signal model is parameterized. 
Figure \ref{fig:eff} shows the product of the efficiency and acceptance as a function of $\ma$ for the intermediate mass hypotheses. 
The BDT selection efficiency is higher in the regions with lower diphoton background contributions because 
the diphoton invariant mass related information ($(\ma-\mhyp)/\mllgg$) is included in the BDT training. 
Since the diphoton background contamination is minimal for masses around 5\GeV, the product of the efficiency and acceptance has a maximum at $\ma\simeq5\GeV$. 
The photons from the ALP decay are easier to distinguish in the high mass regions, which makes the separation between the signal and background by the BDT easier. 
This effect leads to an increase of the product of the efficiency and acceptance beyond 20\GeV.
A better momentum resolution makes the efficiency times acceptance of the muon channel higher than that of the electron channel. 
However, the efficiency decreases because of higher pileup in later data-taking years.
For each intermediate point, a signal model is constructed using the \mllgg shape of the nearest mass hypothesis 
and the normalization is interpolated from the two nearest mass hypotheses.

\section{Background model}\label{sec:bkgmodel}

\begin{figure}[hbt!]
    \centering
      \includegraphics[width=0.49\textwidth]{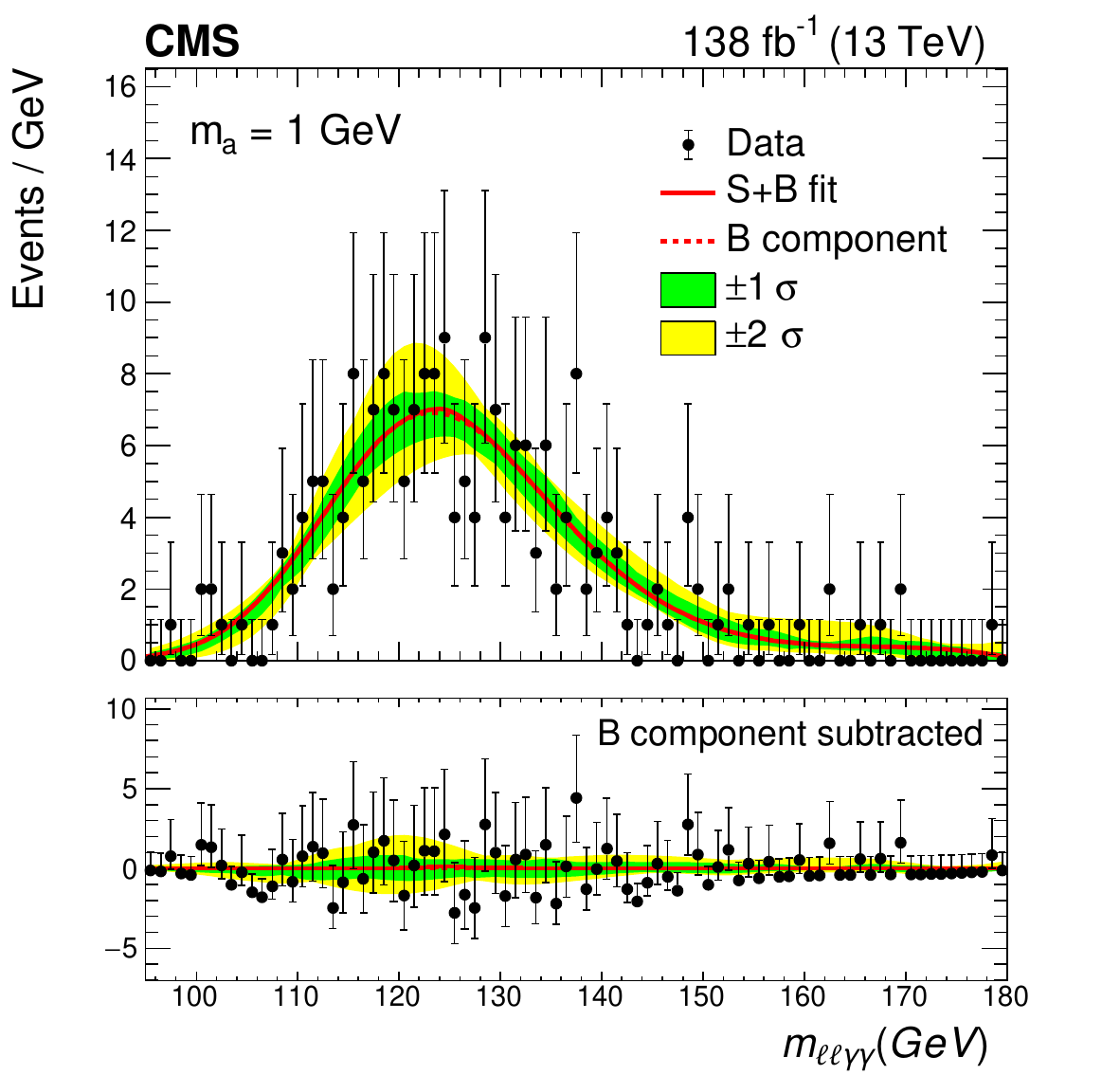}
      \includegraphics[width=0.49\textwidth]{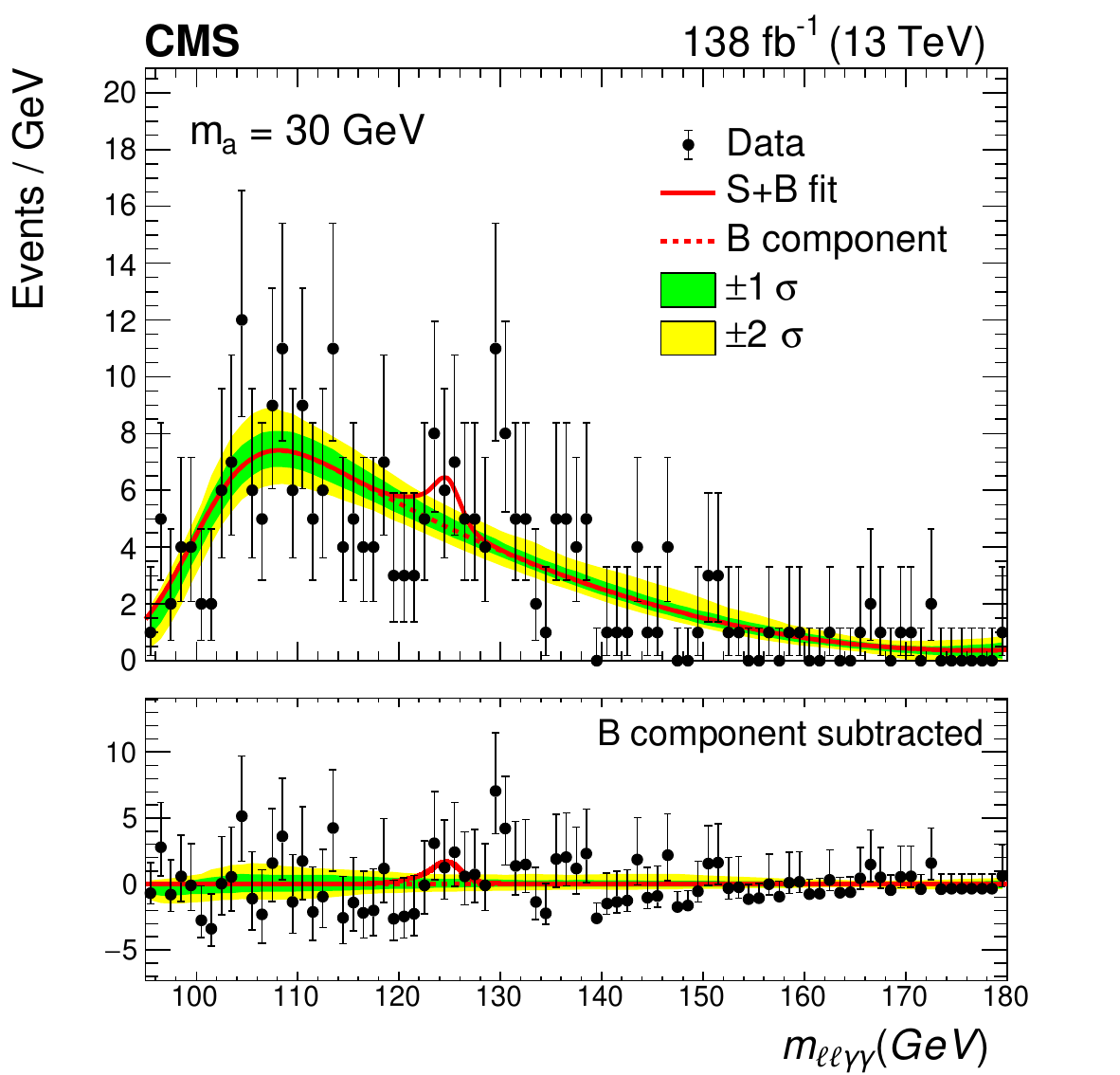}
      \caption{Invariant mass $\mllgg$ distribution in data (black points). The signal-plus-background model fit is shown for $\ma = 1$ (\cmsLeft) and 30 (\cmsRight)\GeV, where the solid red line shows the total signal-plus-background contribution, and the dashed red line shows the background component only. The lower panels show the residual signal yield after the background subtraction. The one (green, inner) and two (yellow, outer) standard deviation bands show the uncertainties in the fitted background model. These bands include the uncertainty due to the choice of function and the uncertainty in the fitted parameters.}
      \label{fig:best_bkg}
\end{figure}

The background model is built to describe the shape of the $\mllgg$ distribution that comes from processes other than the decay of the \PH boson. 
Since the shape of this distribution is not known, different functional forms must be considered. They result in different 
numbers of estimated background events in the signal region, affecting the signal extraction. 
The uncertainty arising from the choice of background model is handled with the envelope method~\cite{DiscreteProfilingMethod}: 
the background modeling is performed using data as in Ref.~\cite{Khachatryan:2014ira}, and the choice of the background functional form 
is treated as a discrete nuisance parameter in the likelihood fit to data.
As shown in Fig.~\ref{fig:best_bkg}, the $\mllgg$ distribution consists of a turn-on peak around 105--115\GeV, 
driven by the photon \pt selection, 
and a falling spectrum at higher mass. To model the background distribution, functions with the following general form are considered:
\ifthenelse{\boolean{cms@external}}{
\begin{multline}
  \mathcal{F}(\mllgg;\mu,\sigma,s,\vec{\alpha}) \\ = \int\mathcal{N}(\mu,\sigma)(\mllgg-t)
  \, f(t;\vec{\alpha})\Theta(s,t)\rd{t},
\end{multline}
}
{
\begin{equation}
  \mathcal{F}(\mllgg;\mu,\sigma,s,\vec{\alpha}) = \int\mathcal{N}(\mu,\sigma)(\mllgg-t)
  f(t;\vec{\alpha})\Theta(s,t)\,\rd{t},
\end{equation}  
}
where $\mathcal{N}$ is a Gaussian function with mean $\mu$ and standard deviation $\sigma$, $\Theta$ is the Heaviside step function defining the cutoff point $s$ below which the falling spectrum function $f$ with shape parameters $\vec{\alpha}$ has no influence, 
and $f$ is a falling spectrum function with shape parameters $\vec{\alpha}$. 
The falling spectrum function families considered in the analysis are exponentials, Bernstein polynomials, Laurent series, and power-law functions. 
A subset of functions from each family is used to build the background model. For each family, 
the maximum order of parameters to be used is chosen with an F-test, 
and the minimum order is determined by applying a requirement on the goodness-of-fit to the data.
A penalty is added to take into account the number of floating parameters in each candidate function.
When making a measurement of a given parameter of interest, 
the discrete profiling method minimizes the overall negative logarithm of the likelihood considering all allowed functions.

The fit is performed over the range $95 < \mllgg < 180\GeV$ where the background model is extended from the sideband region ($95 < \mllgg < 115\GeV$ and $135 < \mllgg < 180\GeV$) to the signal region, and data from all data-taking years and from the electron and muon channels 
are combined to construct the background model. 
A unique background model is created for each \ma hypothesis.
For intermediate \ma hypotheses, the BDT boundary is taken from the closest simulated \ma.
Figure~\ref{fig:best_bkg} shows the best-fit functions for $\ma=1$ and 30\GeV.

For each \ma hypothesis, an ensemble of pseudo-experiments was generated using the various background functions. 
Each pseudo-experiment was fitted using the discrete profiling method, and it was verified that the chosen functional form 
used to describe the background does not introduce any bias in the signal extraction.

\section{Systematic uncertainties}\label{sec:syst}

The systematic uncertainty associated with the background modeling is taken into account with the envelope method, 
as described in Section~\ref{sec:bkgmodel}. The signal modeling has systematic uncertainties of two kinds: 
those that modify the shape of the $\mllgg$ distribution, 
and those that leave the shape of $\mllgg$ distribution unchanged but affect the overall normalization of the signal process.

Uncertainties affecting the shape of the $\mllgg$ distribution are typically related to the energy of the individual photons and leptons, 
and therefore impact the mean and width of the signal model. Their sources are the following: 

\begin{itemize}
    \item Photon energy scale and resolution: corrections are applied to the photon energy scale in data and to the energy resolution in simulation. The uncertainties related to these corrections are computed using $\PZ\to\Pe\Pe$ events~\cite{CMS:2020uim}. The resulting uncertainty on the signal strength due to the uncertainty on the energy scale and resolution is estimated to be 5.7\% (2016), 3.5\% (2017), and 4.5\% (2018). The uncertainties for each data set are fully correlated;
    \item Lepton energy scale and resolution: corrections are applied to the lepton energy scale in data and to the energy resolution in simulation. The uncertainties related to these corrections are computed using $\PZ\to\Pell\Pell$ events~\cite{Sirunyan:2018fpa,CMS:2020uim}. The resulting uncertainty on the signal strength due to the uncertainty on the energy scale and resolution is estimated to be 4.3 (4.9)\% (2016), 4.2 (4.4)\% (2017), and 4.9 (5.2)\% (2018) in the electron (muon) channel. The uncertainties for each data set are fully correlated.
\end{itemize}

The uncertainties that affect the normalization of the signal model are the following: 

\begin{itemize}
    \item Integrated luminosity: uncertainties in the integrated luminosity measurement are estimated to be 1.2\% (2016), 2.3\% (2017), and 2.5\% (2018)~\cite{CMS-PAS-LUM-17-001,CMS-PAS-LUM-17-004,CMS-PAS-LUM-18-002}. The uncertainty in the total integrated luminosity of the three years combined is 1.6\%. The uncertainties for each data set are partially correlated and account for the common sources in the luminosity measurement schemes;
    \item Pileup modeling: the total inelastic $\Pp\Pp$ cross section is varied by $\pm 5\%$, and the analysis is repeated with the shifted weights. Then, the maximum difference of the yield compared to the nominal yield is taken as the systematic uncertainty. The average magnitude of the resulting uncertainty is below 3\% across the full $\ma$ range. The uncertainties for each data set are fully correlated; 
    \item Lepton and photon identification efficiencies: the analysis is rerun by shifting the lepton and photon identification scale factors, which are applied to simulations to match the lepton and photon selection efficiencies in data, by one standard deviation and the maximum difference of the yield with respect to the nominal yield is taken as the systematic uncertainty. The average magnitude of the resulting uncertainty is around 10\% across the full $\ma$ range. The uncertainties for each data set are fully correlated;
    \item The BDT uncertainties: because the BDT score is used to define the signal region, uncertainties in the BDT can lead to event migration across boundaries. The systematic uncertainties in the input variables, which include the data-MC differences in the shower shape variables, are propagated to the final BDT selection. Using the same BDT score boundaries to define the signal region, the maximum difference of the yield with respect to the nominal yield is taken as the systematic uncertainty. The resulting uncertainty is less than 2\%.
\end{itemize}

Theoretical PDF and scale uncertainties have a negligible impact on the analysis results and therefore are not applied. A summary of all the systematic uncertainties is given in Table~\ref{tab:systematics}. The impact of systematic uncertainties on the expected upper limit in the cross section is about 1\% across the probed $\ma$ range and the analysis is primarily limited by the statistical uncertainty.

\begin{table} [hbtp]
    \centering
    \topcaption{Sources of systematic uncertainties and their impact on the signal strength for each data-taking period.}
    \cmsTable{
        \begin{tabular}{lccc}
        $\mllgg$ distribution shape & 2016 & 2017 & 2018  \\
        \hline
        Photon energy scale & ${<}0.1\%$ & ${<}0.1\%$ & ${<}0.1\%$ \\
        Photon energy resolution & 5.7\% & 3.5\% & 4.5\% \\
        Electron energy scale & ${<}0.1\%$ & ${<}0.1\%$ & ${<}0.1\%$ \\ 
        Electron energy resolution & 4.3\% & 4.2\% & 4.9\% \\ 
        Muon energy scale & ${<}0.1\%$ & ${<}0.1\%$ & ${<}0.1\%$ \\
        Muon energy resolution  & 4.9\% & 4.4\% & 5.2\% \\ 
        \hline
        Signal model normalization &  &  &  \\
        \hline
        Integrated luminosity & 1.2\% & 2.3\% & 2.5\% \\
        Pileup modeling & 2.9\% & 2.9\% & 2.5\% \\
        Photon efficiency & 10\% & 10\% & 10\% \\
        Electron efficiency & 1.7\% & 1.5\% & 1.6\% \\
        Muon efficiency & 0.8\% & 0.5\% & 0.5\% \\
        BDT uncertainties & ${<}2\%$ & ${<}2\%$ & ${<}2\%$ \\
        \end{tabular}
    \label{tab:systematics}}
\end{table}

\section{Results and interpretation}\label{sec:results}

\begin{figure}[hbtp]
  \centering
    \includegraphics[width=0.49\textwidth]{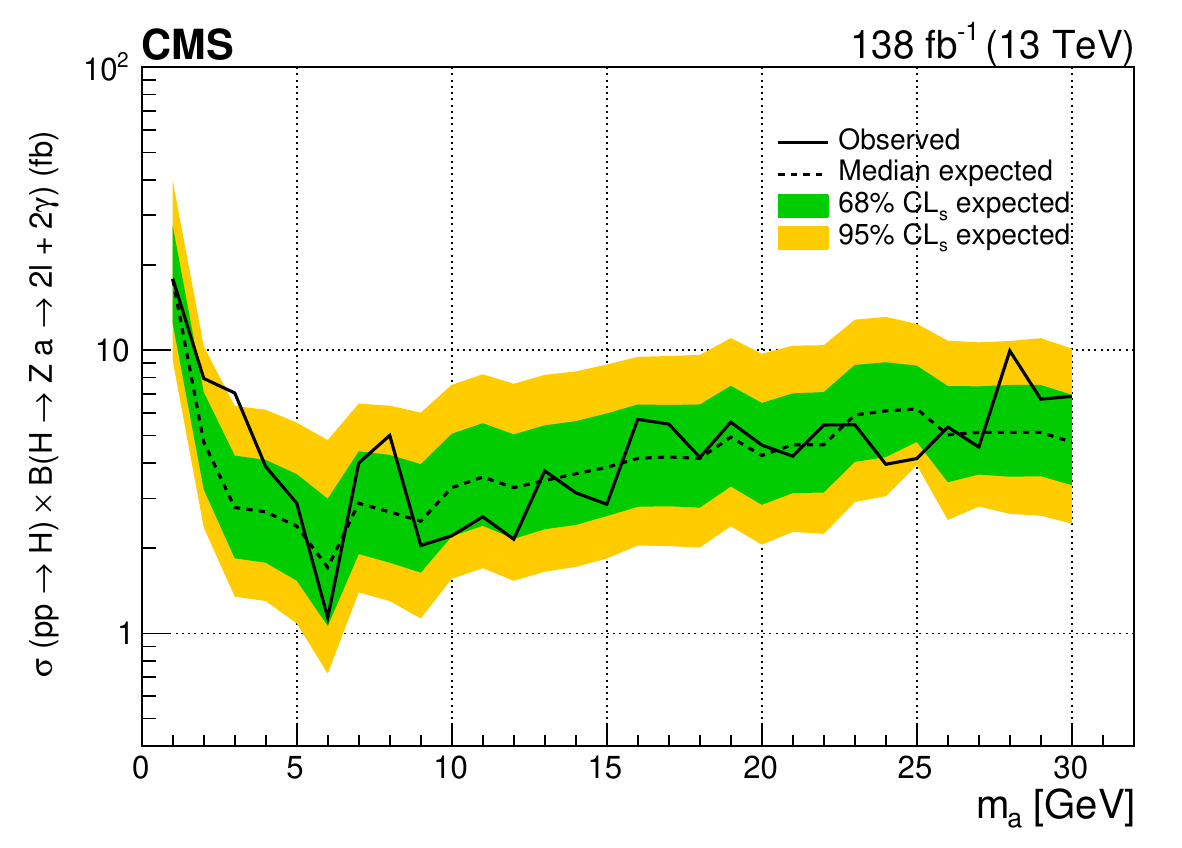}
    \caption{Expected and observed 95\% \CL limits on the product of the production cross section of the \PH boson and its branching fraction into a dilepton and a diphoton pair via a \PZ boson and a pseudoscalar, $\sigma (\Pp\Pp \to \PH)\, \mathcal{B}(\PH \to \PZ \Pa \to \Pell\Pell\PGg\PGg)$. The dashed black curve is the expected upper limit, while the one and two standard-deviation bands are shown in green and yellow, respectively. The solid black curve is the observed upper limit.}
    \label{fig:limit}
\end{figure}

Upper limits at 95\% confidence level (\CL) on the cross section times branching fraction of the \PH boson decaying to a \PZ boson and an ALP 
are evaluated using a \CLs criterion, taking the profile likelihood as a test statistic~\cite{cite:l1,cite:l2,cite:l3}. 
In this procedure, an asymptotic approximation for the likelihood was used~\cite{Cowan:2010js}.
The limits are calculated in the mass range of 1--30\GeV and are shown in Fig.~\ref{fig:limit}.
The observed (expected) limits are in the range 1.1--17.8 (1.7--17.9)\unit{fb} within the probed $\ma$ interval of 1--30\GeV.
An excess of data above the expected SM background with 2.6 (1.3) $\sigma$ local (global) significance~\cite{lookelsewhere} is observed for a mass hypothesis of $\ma=3\GeV$.

\begin{figure}[hbtp]
  \centering
    \includegraphics[width=0.49\textwidth]{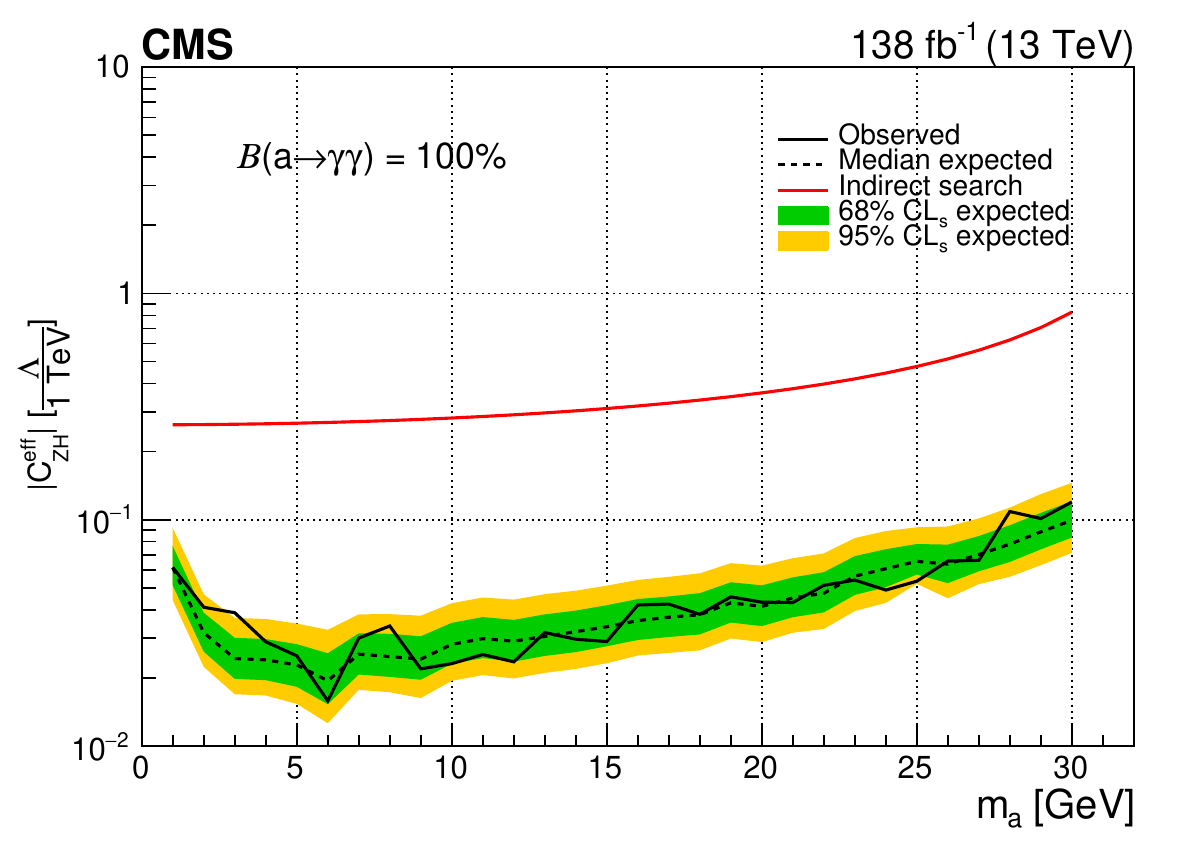}
    \caption{Expected and observed limits at 95\% \CL on \cZh, assuming the ALP decays exclusively to a photon pair. The dashed black curve is the expected upper limit, while the one and two standard-deviation bands are shown in green and yellow, respectively. The solid black curve is the observed upper limit. The red curve represents the interpreted results from the search for invisible decays of the Higgs boson performed by the CMS Collaboration.}
    \label{fig:limit_cZH}
\end{figure}

Upper limits at 95\% \CL are calculated on \cZh~\cite{Georgi:1986df,Bauer:2017ris}, 
as shown in Fig.~\ref{fig:limit_cZH}, where $C_{\PZ\PH}^\text{eff}$ is the effective coupling parameter of the \PH boson, 
the \PZ boson, and the ALP, and $\Lambda$ is the scale of BSM physics. 
In this interpretation, in order to avoid the issue of ALPs being long-lived in the detector, the ALP is assumed to decay promptly with the branching fraction $\mathcal{B}(\Pa\to\PGg\PGg) = 100\%$~\cite{CMS:2021pcy}.
The indirect limit from the search for invisible decays of the Higgs boson performed by the CMS Collaboration~\cite{CMS:2023sdw} is also shown.
Whereas the expected limits on the \PH boson production cross section times its branching fraction into a dilepton and a diphoton pair via a \PZ boson and a pseudoscalar, 
shown in Fig.~\ref{fig:limit}, do not exhibit a strong mass dependence for $\ma>3\GeV$, the limits on the \cZh are several factors stronger in the low-mass region with respect to $\ma=30\GeV$.

\section{Summary}\label{sec:summary}

A search for Higgs boson (\PH) decays to a \PZ boson and an axion-like particle (ALP), which subsequently decay into a lepton pair and a photon pair, respectively, 
is presented. The analysis is based on proton-proton collision data collected at $\sqrt{s} = 13$\TeV by the CMS experiment 
in 2016--2018, corresponding to an integrated luminosity of 138\fbinv. The analysis probes pseudoscalar masses ($\ma$) in the range 
1--30\GeV. This is the first search for Higgs boson decays in the final state of two leptons and two photons. 
Upper limits are set at 95\% confidence level on the production cross section of the Higgs boson times its 
branching fraction into a dilepton and a diphoton pair via a \PZ boson and a pseudoscalar, $\sigma (\Pp\Pp\to \PH)\, \mathcal{B}(\PH\to\PZ \Pa \to \Pell\Pell\PGg\PGg)$, where $\Pell = \Pe, \PGm$. 
The observed (expected) limits varies in the range 1.1--17.8 (1.7--17.9)\unit{fb} within the probed $\ma$ interval of 1--30 GeV. 
The largest excess with respect to the standard model prediction is observed for $\ma=3\GeV$ and has a local (global) significance of 2.6 (1.3) standard deviations.
Constraints are set on the ALP model parameter \cZh, which describes 
the coupling between the Higgs boson, \PZ boson, and ALP. 

\begin{acknowledgments}
We congratulate our colleagues in the CERN accelerator departments for the excellent performance of the LHC and thank the technical and administrative staffs at CERN and at other CMS institutes for their contributions to the success of the CMS effort. In addition, we gratefully acknowledge the computing centers and personnel of the Worldwide LHC Computing Grid and other centers for delivering so effectively the computing infrastructure essential to our analyses. Finally, we acknowledge the enduring support for the construction and operation of the LHC, the CMS detector, and the supporting computing infrastructure provided by the following funding agencies: SC (Armenia), BMBWF and FWF (Austria); FNRS and FWO (Belgium); CNPq, CAPES, FAPERJ, FAPERGS, and FAPESP (Brazil); MES and BNSF (Bulgaria); CERN; CAS, MoST, and NSFC (China); MINCIENCIAS (Colombia); MSES and CSF (Croatia); RIF (Cyprus); SENESCYT (Ecuador); MoER, ERC PUT and ERDF (Estonia); Academy of Finland, MEC, and HIP (Finland); CEA and CNRS/IN2P3 (France); SRNSF (Georgia); BMBF, DFG, and HGF (Germany); GSRI (Greece); NKFIH (Hungary); DAE and DST (India); IPM (Iran); SFI (Ireland); INFN (Italy); MSIP and NRF (Republic of Korea); MES (Latvia); LAS (Lithuania); MOE and UM (Malaysia); BUAP, CINVESTAV, CONACYT, LNS, SEP, and UASLP-FAI (Mexico); MOS (Montenegro); MBIE (New Zealand); PAEC (Pakistan); MES and NSC (Poland); FCT (Portugal); MESTD (Serbia); MCIN/AEI and PCTI (Spain); MOSTR (Sri Lanka); Swiss Funding Agencies (Switzerland); MST (Taipei); MHESI and NSTDA (Thailand); TUBITAK and TENMAK (Turkey); NASU (Ukraine); STFC (United Kingdom); DOE and NSF (USA).

\hyphenation{Rachada-pisek} Individuals have received support from the Marie-Curie program and the European Research Council and Horizon 2020 Grant, contract Nos.\ 675440, 724704, 752730, 758316, 765710, 824093, and COST Action CA16108 (European Union); the Leventis Foundation; the Alfred P.\ Sloan Foundation; the Alexander von Humboldt Foundation; the Science Committee, project no. 22rl-037 (Armenia); the Belgian Federal Science Policy Office; the Fonds pour la Formation \`a la Recherche dans l'Industrie et dans l'Agriculture (FRIA-Belgium); the Agentschap voor Innovatie door Wetenschap en Technologie (IWT-Belgium); the F.R.S.-FNRS and FWO (Belgium) under the ``Excellence of Science -- EOS" -- be.h project n.\ 30820817; the Beijing Municipal Science \& Technology Commission, No. Z191100007219010 and Fundamental Research Funds for the Central Universities (China); the Ministry of Education, Youth and Sports (MEYS) of the Czech Republic; the Shota Rustaveli National Science Foundation, grant FR-22-985 (Georgia); the Deutsche Forschungsgemeinschaft (DFG), under Germany's Excellence Strategy -- EXC 2121 ``Quantum Universe" -- 390833306, and under project number 400140256 - GRK2497; the Hellenic Foundation for Research and Innovation (HFRI), Project Number 2288 (Greece); the Hungarian Academy of Sciences, the New National Excellence Program - \'UNKP, the NKFIH research grants K 124845, K 124850, K 128713, K 128786, K 129058, K 131991, K 133046, K 138136, K 143460, K 143477, 2020-2.2.1-ED-2021-00181, and TKP2021-NKTA-64 (Hungary); the Council of Science and Industrial Research, India; ICSC -- National Research Center for High Performance Computing, Big Data and Quantum Computing, funded by the EU NexGeneration program (Italy); the Latvian Council of Science; the Ministry of Education and Science, project no. 2022/WK/14, and the National Science Center, contracts Opus 2021/41/B/ST2/01369 and 2021/43/B/ST2/01552 (Poland); the Funda\c{c}\~ao para a Ci\^encia e a Tecnologia, grant CEECIND/01334/2018 (Portugal); the National Priorities Research Program by Qatar National Research Fund; MCIN/AEI/10.13039/501100011033, ERDF ``a way of making Europe", and the Programa Estatal de Fomento de la Investigaci{\'o}n Cient{\'i}fica y T{\'e}cnica de Excelencia Mar\'{\i}a de Maeztu, grant MDM-2017-0765 and Programa Severo Ochoa del Principado de Asturias (Spain); the Chulalongkorn Academic into Its 2nd Century Project Advancement Project, and the National Science, Research and Innovation Fund via the Program Management Unit for Human Resources \& Institutional Development, Research and Innovation, grant B05F650021 (Thailand); the Kavli Foundation; the Nvidia Corporation; the SuperMicro Corporation; the Welch Foundation, contract C-1845; and the Weston Havens Foundation (USA).
\end{acknowledgments}

\bibliography{auto_generated}

\cleardoublepage \appendix\section{The CMS Collaboration \label{app:collab}}\begin{sloppypar}\hyphenpenalty=5000\widowpenalty=500\clubpenalty=5000
\cmsinstitute{Yerevan Physics Institute, Yerevan, Armenia}
{\tolerance=6000
A.~Hayrapetyan, A.~Tumasyan\cmsAuthorMark{1}\cmsorcid{0009-0000-0684-6742}
\par}
\cmsinstitute{Institut f\"{u}r Hochenergiephysik, Vienna, Austria}
{\tolerance=6000
W.~Adam\cmsorcid{0000-0001-9099-4341}, J.W.~Andrejkovic, T.~Bergauer\cmsorcid{0000-0002-5786-0293}, S.~Chatterjee\cmsorcid{0000-0003-2660-0349}, K.~Damanakis\cmsorcid{0000-0001-5389-2872}, M.~Dragicevic\cmsorcid{0000-0003-1967-6783}, A.~Escalante~Del~Valle\cmsorcid{0000-0002-9702-6359}, P.S.~Hussain\cmsorcid{0000-0002-4825-5278}, M.~Jeitler\cmsAuthorMark{2}\cmsorcid{0000-0002-5141-9560}, N.~Krammer\cmsorcid{0000-0002-0548-0985}, D.~Liko\cmsorcid{0000-0002-3380-473X}, I.~Mikulec\cmsorcid{0000-0003-0385-2746}, J.~Schieck\cmsAuthorMark{2}\cmsorcid{0000-0002-1058-8093}, R.~Sch\"{o}fbeck\cmsorcid{0000-0002-2332-8784}, D.~Schwarz\cmsorcid{0000-0002-3821-7331}, M.~Sonawane\cmsorcid{0000-0003-0510-7010}, S.~Templ\cmsorcid{0000-0003-3137-5692}, W.~Waltenberger\cmsorcid{0000-0002-6215-7228}, C.-E.~Wulz\cmsAuthorMark{2}\cmsorcid{0000-0001-9226-5812}
% ...
\par}
\vskip\cmsinstskip
\dag:~Deceased\\
$^{1}$Also at Yerevan State University, Yerevan, Armenia\\
$^{2}$Also at TU Wien, Vienna, Austria\\
% ...
\end{sloppypar}
\end{document}